\pdfoutput=1
\documentclass[journal, onecolumn,12pt]{IEEEtran}
\usepackage[margin=1in]{geometry}
\usepackage{setspace}   

\usepackage{pgf, tikz}
\usepackage{calc}
\usetikzlibrary{calc}
\usetikzlibrary{arrows}
\definecolor{mycolor1}{rgb}{0.10000,0.60000,0.30000}%
\usepackage{comment}
\usetikzlibrary{decorations,calligraphy}
\usepackage{cancel}
\usepackage[normalem]{ulem}
\usepackage{xcolor}

\definecolor{mycolor1}{rgb}{0.10000,0.60000,0.30000}%
\definecolor{mypink2}{RGB}{219, 48, 122}
\usepackage{amssymb}
\usepackage{amsmath}
\usepackage{color}
\usepackage{bbm}
\usepackage{research5IEEE}

\newcommand{\defeq}{\triangleq}

\newfont{\bbb}{msbm10 scaled 500}

\newfont{\bb}{msbm10 scaled 1100}

\newcommand{\CDc}{{\cal{CD}}}
\newcommand{\Cc}{{\cal C}}
\newcommand{\Dc}{{\cal D}}

\newcommand{\Sc}{{\cal S}}
\newcommand{\Tc}{{\cal T}}
\newcommand{\Uc}{{\cal U}}
\newcommand{\Wc}{{\cal W}}

\newcommand{\Xc}{{\cal X}}
\newcommand{\Yc}{{\cal Y}}
\newcommand{\Zc}{{\cal Z}}

\usepackage{color}
\usepackage{cite}
\usepackage{graphics}
\usepackage{tikz}
\usepackage{pgfplots}
\usepackage{stfloats}

\usepackage{float}
\usepackage{placeins}
\usetikzlibrary{shapes,  arrows,  decorations.markings,  arrows.meta}
\usetikzlibrary{patterns} 
\allowdisplaybreaks
\usepackage{epsfig}

\usepackage{rotating}
\usepackage{amsmath}
\usepackage{amsfonts}
\usepackage{url}

\usepackage[caption=false, font=footnotesize]{subfig}
\usepackage{epstopdf}
\usepackage{amsthm}
\usepackage{mathtools}

\usepackage{algorithm}
\usepackage[noend]{algpseudocode}

\usepackage{array}

\usepackage{cite}

\newtheorem{theorem}{Theorem}
\newtheorem{example}{Example}
\newtheorem{definition}{Definition}

\newtheorem{corollary}{Corollary}
\newtheorem{remark}{Remark}
\newcommand{\mw}[1]{{\color{black}#1}}
\definecolor{darkgreen}{RGB}{0,128,0}
\definecolor{blue-green}{rgb}{0, 0.6, 0.8}

\newcommand{\D}{\mathsf{D}}

\definecolor{dgreen}{rgb}{0, 0.8, 0.4}
\newcommand{\lo}[1]{{\color{black}#1}}
\newcommand{\sh}[1]{}

\newcommand{\R}{\mathsf{R}}

\newcommand*{\colorboxed}{}
\def\colorboxed#1#{%
	\colorboxedAux{#1}%
}
\newcommand*{\colorboxedAux}[3]{%
	\begingroup
	\colorlet{cb@saved}{.}%
	\color#1{#2}%
	\boxed{%
		\color{cb@saved}%
		#3%
	}%
	\endgroup
}

\allowdisplaybreaks[4]
\pgfplotsset{compat=1.16} 
\title{An Information-Theoretic Approach to Collaborative Integrated Sensing and Communication for Two-Transmitter Systems }
\begin{document}
	\doublespacing
	\author{
		\IEEEauthorblockN{Mehrasa Ahmadipour and  Mich\`ele Wigger} \\ 
		\IEEEauthorblockA{\small\IEEEauthorrefmark{1} LTCI Telecom Paris, IP Paris, 91120 Palaiseau, France, Emails:
			\url{{mehrasa.ahmadipour,michele.wigger}@telecom-paris.fr}\\}
	}
	\maketitle
	\vspace{-2cm}
	\begin{abstract} 
		This paper considers information-theoretic models for integrated sensing and communication (ISAC) over   multi-access channels (MAC) and device-to-device (D2D) communication. The models are  general and  include as special cases scenarios with and without perfect or imperfect state-information at the MAC receiver as well as causal state-information at the D2D terminals. For both setups, we propose collaborative sensing ISAC schemes where terminals not only convey data to the other terminals but also state-information that they  extract from their previous observations. This state-information can be exploited at the other terminals to improve their sensing performances. Indeed, as we show through examples, our schemes improve over previous non-collaborative schemes in terms of their achievable rate-distortion tradeoffs. For D2D we propose two schemes, one where  compression of state information is separated from channel coding and one where it is integrated via a hybrid coding approach. 
		
	\end{abstract}
	
	\IEEEpeerreviewmaketitle
	\vspace{-0cm}
	\section{Introduction}

	Next-generation wireless networks are expected to support several autonomous and intelligent applications that rely heavily on accurate sensing and localization techniques \cite{bourdoux20206g}. Important  examples are intelligent transport systems, where vehicles continuously sense environmental changes and simultaneously exchange sensing-information  and data with  already detected vehicles, base stations, or central servers. Such  simultaneous sensing and data-communication applications are also the focus of this work. More specifically, we are interested in multi-terminal scenarios where different terminals communicate data with each other and simultaneously exploit the backscattered signals for sensing purposes.

	
	A common but naive approach to address sensing and communication is to separate the two tasks in independent systems and split the available resources such as bandwidth and power between the two systems. In our information-theoretic model, such a system
	corresponds to resource-sharing (e.g., time-sharing) between communication and sensing.
	However, the high cost of spectrum and hardware encourages integrating the sensing and communications tasks
	via a single waveform and a single hardware platform \cite{zheng2019radar,liu2020joint}.
	\textcolor{black}{A large body of works studied  integrated sensing and communication (ISAC) scenarios from a communication-theoretic or signal-processing perspective (see, e.g.,  \cite{sturm2011waveform,gaudio2019effectiveness} and references therein)}, mostly investigating appropriate choices for the employed   waveform that in ISAC applications has to serve both the  sensing and the communication tasks. Interestingly, different tradeoffs between the communication and sensing performances can be obtained by changing the employed waveform.

	The fundamental performance limits of integrated sensing and communication systems were first considered in
	\cite{kobayashi2018joint}. Specifically, \cite{kobayashi2018joint} introduced an information-theoretic model for integrated sensing and communication based on a generalized-feedback model, which  captures two underlying assumptions used in radar signal processing.
	On the one hand, generalized feedback captures the inherently passive nature of the backscattered signal observed at the transmitter (Tx), which cannot be controlled but is determined by its surrounding environment. On the other hand, it models the fact that the backscattered signal depends on the waveform employed by the Tx. It was proposed to use the classical average per-letter block-distortion to measure the Tx's sensing performance on the i.i.d. state-sequence. The authors of \cite{kobayashi2018joint}, see also \cite{Ahmadipour2022IT}   characterized the exact \emph{capacity-distortion tradeoff} of arbitrary discrete memoryless channels (DMCs) with generalized feedback. This quantity naturally measures the inherent tradeoff between increasing data rate and reducing  sensing distortion in such integrated systems. 
	Interestingly, the results show that the optimal tradeoff is achieved by standard random code constructions as used for traditional data communication, where  the statistics of the channel inputs (and thus of the codewords)  however has to be adapted to meet  the desired sensing performance. Notice that this observation is consistent with the signal-processing literature on the search for adequate channel input waveforms which allow to meet the desired sensing performance while still achieving high communication rates. Similar results were also derived for discrete memoryless broadcast channels (DMBCs) \cite{Ahmadipour2022IT} where a single transmitter communicates with two receivers. Both the DMC and the DMBC are thus single-Tx networks, and the optimal sensing is a simple per-symbol estimation of the hidden state given the  channel inputs and outputs at the sensing terminal. The optimality of such a  simple symbol-by-symbol estimator stems from the fact that for a fixed input sequence the generalized feedback channels and the state-sequence both behave in a memoryless manner. 
	
	The sensing situation becomes  more interesting and challenging when the sensing terminal is not the only terminal feeding inputs to the channel. In this case, the  effective disturbance for the sensing is not necessarily memoryless since the inputs from the other terminals also create disturbances and can have memory. In this case, a strategy that first attempts to guess the other Txs' codewords  followed by a symbol-wise estimator based on the observations and the guessed codewords can lead  to a smaller (and thus better) distortion. This has also  been observed in   \cite{choudhuri2013causal}, where communication is over a DMC and state estimation is performed at the receiver (Rx) side. In this case, the optimal sensing strategy is first to decode the Tx's codeword and then apply an optimal symbol-by-symbol estimator to this codeword and  the observed channel outputs. A similar strategy  was applied in the two-transmitter single-Rx multi-access channel (MAC) ISAC scenario of \cite{kobayashi2019joint} where through the generalized feedback each Tx first decodes part of the data sent by the other Tx and then applies a symbol-by-symbol estimator to the decoded codeword as well as its own channel inputs and outputs.  In fact, the ISAC scheme   of \cite{kobayashi2019joint} is based on Willems' scheme for the MAC with generalized feedback, where each Tx encodes its data into two super-positioned codewords, whereof the lower data-layer is decoded by the other Tx. This  data is then repeated by both Txs  in the next block as part of a third lowest-layer codeword, allowing the two Txs to transmit data cooperatively  Somewhat naturally, \cite{kobayashi2019joint} suggests to use this decoded lower  data-layer also for sensing purposes in the sense that each Tx applies  the symbol-by-symbol estimator not only to its inputs and outputs but also to this decoded codeword. In this article, which is based on the conference paper \cite{Mehrasa2022MACISIT}, we suggest to use this decoded codeword not only to exchange data,
	but also to exchange sensing information. The  concept of exchanging sensing information for ISAC has  been studied in the signal processing literature under the paradigm of \emph{collaborative sensing}. 
	
	In this sense, we  introduce the concept of  collaborative sensing  for ISAC also to the information-theoretic literature, where we focus on the MAC and the related device-to-device (D2D) communication, i.e., the two-way channel. For the MAC, we naturally extend Willem's coding scheme so as to convey also state-information from one Tx to the other over the communication path that is built over the generalized feedback link. The proposed scheme can be considered as a separate source-channel coding scheme in the sense that each Tx first compresses the obtained outputs and inputs so as to extract state information, and then transmits the compression index using a pure channel code (here Willems' coding scheme) to the other Tx. The proposed scheme obtains a better sensing performance than a previous ISAC scheme \cite{kobayashi2019joint} without collaborative sensing, and thus a better distortion-capacity tradeoff. For D2D communication, we present a similar collaborative sensing ISAC scheme based on source-channel separation and using Han's two-way channel scheme. Furthermore, we present an improved scheme that is based on joint source-channel coding (JSCC), more specifically on hybrid coding. We show enhanced performances  of both simple collaborative sensing schemes. In both the MAC and the D2D scenario, \textcolor{black}{the maximum rates  achieved by our proposed scheme for given sensing distortions are strictly concave functions of the distortion pairs,} and thus also improve over classical time- or resource-sharing strategies. 
	
	Recently, various other information-theoretic works have analyzed the fundamental limits of  ISAC systems,  such as \cite{Joudeh2021JBinaryDetect, Joudeh2022Discrim,Bloch2022, Gunlu}. For example, \cite{Gunlu} analyzes systems with secrecy constraints, while 
	\cite{Joudeh2021JBinaryDetect, Joudeh2022Discrim,Bloch2022} study channels that depend on a single fixed  parameter and  transmitters or sensor nodes wish  to estimate this parameter based on backscatter signals. Their model is  thus suited for scenarios where the estimation parameters change at a much slower time scale compared to the channel symbol period. Specifically, while in 
	\cite{Joudeh2022Discrim} sensing (parameter estimation) is performed at the transmitter, in \cite{Joudeh2021JBinaryDetect}  it is performed at a sensor that is close but not collocated with the transmitter. 
	%
	The study in\cite{Bloch2022} analyzes the detection-error exponents of open-loop and close-loop coding strategies. 
	\medskip
	
	\textit{Summary of Contributions and Outline of this Article:} 
	\begin{itemize}
		\item In Section~\ref{sec:MAC_model} we introduce our information-theoretic  ISAC MAC model with state-sensing at the Txs. We also  show that it is of  general nature  and in particular can model scenarios with partial or perfect channel state information at the Rx  as well as scenarios where the Txs wish to reconstruct  functions or distorted versions of the actual state that is governing  the channel. 
		\item  In Section~\ref{sec:MAC_scheme} we describe our collaborative-sensing ISAC MAC scheme and show at hand of examples that it improves both over simple time-sharing as well as over previous schemes. Notice that our scheme does not employ Wyner-Ziv compression, but the equally strong \emph{implicit binning technique}, as used for example  in \cite{Tuncel}. 
		\item Section~\ref{sec:D2D_model}  describes our information-theoretic  ISAC D2D model with state-sensing at both terminals. Again, we show that our model is rather general and includes  scenarios with strictly-causal  perfect or imperfect state-information at the terminals. 
		\item  In Section~\ref{sec:D2D_scheme} we propose two collaborative-sensing ISAC D2D schemes. The first is based on a  separate source-channel coding approach and the second on an improved JSCC approach using hybrid coding. In both schemes, the transmitted codeword carries not only data but also compression information that the other terminal can exploit  for sensing. While the separation-based scheme employs Wyner-Ziv compression to account for the side-information at the other Tx, the JSCC based scheme uses implicity binning as in standard hybrid coding. 
	\end{itemize}
	\medskip
	
	
	\textit{Notations:}
	We use calligraphic letters to denote sets,  e.g.,  $\Xc$. 
	Random variables are denoted by uppercase letters,  e.g.,  $X$,  and their realizations by lowercase letters,  e.g.,  $x$. 	
	For positive integers $n$, we  use $[1:n]$ to denote the set $\{1, \cdots, n\}$,  $X^n$ for the tuple of random variables $(X_1, \cdots, X_n)$ and  $x^n$ for $(x_1,\ldots, x_n)$. 
	We abbreviate \emph{independent and identically distributed} as \emph{i.i.d.} {and \emph{probability mass function} as \emph{pmf}.} Logarithms are taken with respect to base $2$. \textcolor{black}{We shall use $\Tc_{\epsilon}^{N} (P_{XY} )$ to indicate the of strongly jointly-typical sequences $\{(x^n , y^n )\}$ with respect to  the distribution $P_{XY}$ as defined in \cite{csiszar}.} 
	For an index $k\in\{1,2\}$, we define $\bar{k}:=3-k$ and for an event $\mathcal{A}$ we denote its complement by $\bar{\mathcal{A}}$. 
	Moreover, $\mathbbm{1}\{\cdot\}$ denotes the indicator function.
	
	\section{Two-User Multiple-Access Channel with Generalized Feedback: System Model}\label{sec:MAC_model}
	
	In this section we consider the two-user multi-access channel (MAC) with generalized feedback, where two Txs wish to convey independent data to a common Rx and through the generalized feedback link they estimate the respective state sequences $S_1^n$ and $S_2^n$ governing the transition law over the MAC and the  generalized feedback.

	\subsection{System Model}
\begin{figure}[t]
		\centering
		\includegraphics[scale=0.87]{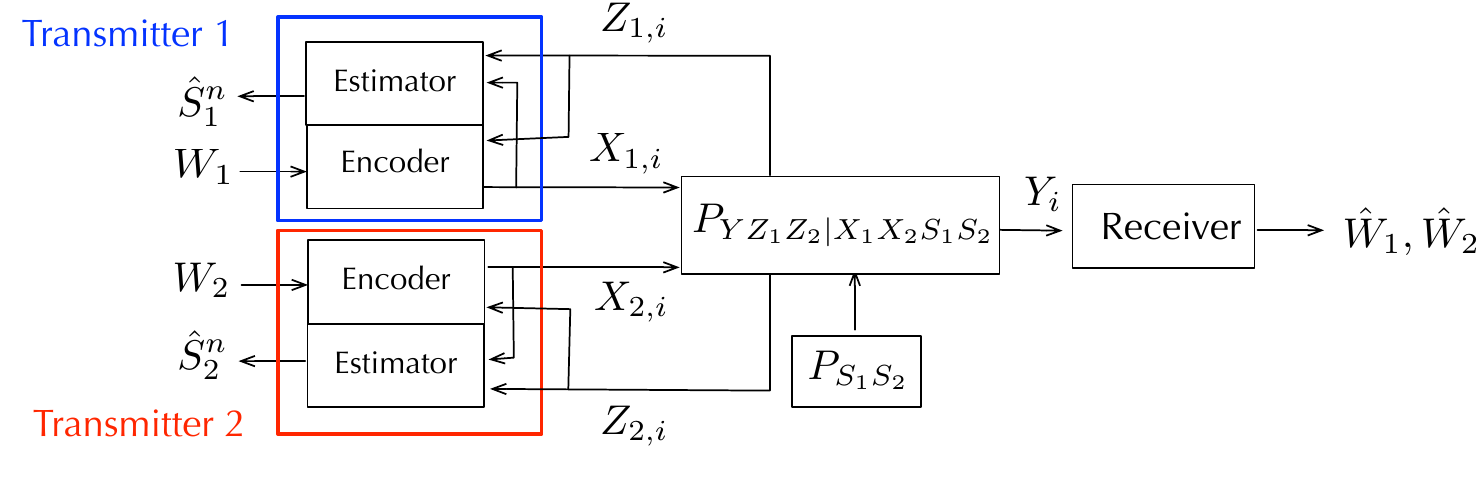}
		
		\caption{State-dependent discrete memoryless multiaccess channel with sensing at the transmitters.}
		\label{fig:ModelMAC}
		
	\end{figure}

	Consider the two-Tx single-Rx MAC scenario  in Fig.~\ref{fig:ModelMAC}.  The model consists of a two-dimensional memoryless state sequence $\{(S_{1, i},  S_{2, i})\}_{i\geq 1}$ whose samples at any given time $i$ are distributed according to a given joint law $P_{S_1S_2}$ over the state alphabets $\Sc_1\times \Sc_2$. 
	Given that at time-$i$ Tx 1 sends input $X_{1,i}=x_1$ and Tx~2 input $X_{2,i}=x_2$ and given  state realizations $S_{1,i}=s_{1}$ and $S_{2,i}=s_{2}$,  the Rx's  time-$i$ output $Y_{i}$  and the Txs'  feedback signals $Z_{1,i}$ and $Z_{2,i}$ are distributed according to the \textcolor{black}{time-invariant} channel transition law $ P_{YZ_1Z_2|S_1S_2X_1X_2}(\cdot,\cdot,\cdot|s_1,s_2,x_1,x_2)$. Input and output 
	alphabets $\Xc_1, \Xc_2,  \Yc,  \Zc_1,  \Zc_2, \Sc_1, \Sc_2$ are assumed  finite.\footnote{\textcolor{black}{Notice that our results  can also be extended to well-behaved continuous channels}.}
	A $(2^{n\R_1}, 2^{n\R_2},  n)$-code  consists  of
	\begin{enumerate}
		\item two message sets $\Wc_1= [1:2^{n\R_1}]$ and $\Wc_2= [1:2^{n\R_2}]$;
		\item a sequence of encoding functions $\Omega_{k,i}\colon \Wc_k \times \Zc_k^{{i-1} }\to \Xc_k$,  for $i=1, 2, \ldots, n$ and $k=1,2$; 
		\item  a decoding function $g \colon \Yc^n \to \Wc_1\times \Wc_2$; 
		\item for each $k=1, 2$  a state estimator  $\phi_k \colon \Xc_k^n \times \Zc_k^n \to \hat{\Sc}_k^n$,  where  $\hat{\Sc}_1$ and $\hat{\Sc}_2$ are given  reconstruction alphabets.
	\end{enumerate}
	%
	%
	%
	\textcolor{black}{Fix a blocklength $n$, rates $R_1, R_2\geq 0$, and a $(2^{nR_1},2^{nR_2},n)$-code $(\{\Omega_{1,i}\},\{\Omega_{2,i}\},  g, \phi_1,\phi_2)$. Let then the random message $W_k$ be uniformly distributed over the message set $\Wc_k$, for each $k=1,2$, and the generate the inputs  according to the encoding function $X_{k,i}=\Omega_{k,i}(W_k, Z_{k}^{i-1})$,  for $i=1, \ldots,  n$.}
	The Txs' state estimates 
	are obtained as $\hat{S}_k^n:= (\hat{S}_{k, 1}, \cdots, \hat{S}_{k, n} )=\phi_k(X_k^n,  Z_k^n)$ and the Rx's guess of the messages as  $(\hat{W}_1, \hat{W}_2)=g(Y^n)$.
	We shall measure the quality of the state estimates $\hat{S}_k^n$   by  bounded per-symbol distortion functions $d_k\colon \Sc_k\times \hat{\Sc}_k \mapsto [0, \infty)$, 
	and consider \emph{expected average block distortions}
	\begin{equation}
		\Delta_k^{(n)}:= \frac{1}{n} \sum_{i=1}^n \mathbb{E}[d_k(S_{k, i},  \hat{S}_{k, i})],  \quad k=1, 2.
	\end{equation}
	The probability of decoding error is defined as:
	\begin{IEEEeqnarray}{rCl}
		P^{(n)}_e& := &\textnormal{Pr}\Big( \hat{W}_1 \neq W_1 
		\quad \textnormal{or} 
		\quad \hat{W}_2\neq W_2 \Big).
	\end{IEEEeqnarray}
	\begin{definition} 
		\hspace{0.5cm}	A rate-distortion tuple $(\R_1,  \R_2,  \D_1,  \D_2)$ is
		achievable if there exists  a sequence (in $n$) of  $(2^{n\R_1}, 2^{n\R_2},  n)$ codes that simultaneously satisfy
		\begin{subequations}\label{MAC:eq:asymptotics}
			\begin{IEEEeqnarray}{rCl}
				\lim_{n\to \infty}	P^{(n)}_e 
				&=&0 \\
				\varlimsup_{n\to \infty}	\Delta_k^{(n)}& \leq& \D_k,  \quad \textnormal{for } k=1, 2.\label{MAC:eq:asymptotics_dis}
			\end{IEEEeqnarray}
		\end{subequations}
	\end{definition}
	\begin{definition}
		The capacity-distortion region $\CDc$ is the closure of the set of all achievable tuples $(\R_1,  \R_2, \D_1, \D_2)$.
	\end{definition}
	
	\vspace{0.3cm}	
	\begin{remark}[On the States]
		Notice  that the general  law $P_{S_1S_2}$ governing the states $S_1^n$ and $S_2^n$ allows to model various types of situations including scenarios where the state sequences are highly correlated (even  identical) or  scenarios where the state-sequences are independent. 
		
		Our model also includes a scenario where the channel is governed by an internal i.i.d. state sequence $S^n$ of pmf $P_S$ and the states $S_1^n, S_2^n$ are related  to $S^n$ over an independent memoryless channel $P_{S_1S_2|S}$. For example, the states $S_1^n$ and $S_2^n$ can be imperfect or noisy versions of the actual state sequence $S^n$. To see that this scenario can be included in our model, notice that since no terminal observes $S^n$ nor attempts to reconstruct $S^n$, both the distortions and the error probabilities only depend on the conditional law 
		\begin{IEEEeqnarray}{rCl}\label{eq:ch}
			P_{YZ_1Z_2|X_1X_2S_1S_2}(y, z_1,z_2|x_1,x_2,s_1,s_2)  &=&
			\nonumber\\&&\hspace{-2cm} \sum_{s} P_{YZ_1Z_2|X_1X_2S}(y, z_1,z_2|x_1,x_2,s) \frac{P_{S}(s) P_{S_1S_2|S}(s_1,s_2|s) }{P_{S_1S_2}(s_1,s_2)},
		\end{IEEEeqnarray}
		where $P_{S_1S_2}(s_1,s_2) = \sum_{s} P_{S}(s) P_{S_1S_2|S}(s_1,s_2|s)$ denotes the joint pmf of the two states.  Computing the channel law in \eqref{eq:ch} and plugging it into our results in the next section, thus immediately also provides results for the described setup where the actual state is $S^n$ and the states $S_1^n$ and $S_2^n$ are noisy versions thereof. 
	\end{remark}

	\begin{remark}[State-Information] \label{rem:state1}Our model also includes scenarios with perfect or imperfect state-information at the Rx. In fact, considering  our model with an output
		\begin{equation}\label{eq:state_info1}
			Y=(T, Y') 
		\end{equation}
		where $Y'$  denotes the actual MAC output and $T$ the Rx's imperfect channel state-information   about the states $S_1^n$ and $S_2^n$ . Notice that in our model, the Rx observes the state-information $T^n$ only in a causal manner. Causality is however irrelevant here since the Rx only has to decode the messages at the end of the entire transmission. 
		Therefore, plugging  the choice \eqref{eq:state_info1} into our results  for  $T$  the Rx state-information and $Y'$ the actual MAC output,  our results in the following section directly lead to results for this related setup with Rx state-information. 	
	\end{remark}
	
	\textcolor{black}{
		\begin{remark}[The Relay-Channel]
			The MAC with generalized-feedback model includes the relay-channel as a special case. It suffices to restrict $R_2=0$, in which case Tx 2 degenerates to a relay terminal. The results we elaborate in the following section does immediately apply also to the relay channel.
	\end{remark}}
	\section{A Collaborative ISAC Scheme for the MAC}\label{sec:MAC_scheme}
	
	Before describing our collaborative ISAC scheme for the MAC, we review literature on the MAC  and in particular  Willem's scheme for the MAC with generalized feedback, which acts as a building block \textcolor{black}{for} our scheme. 
	
	While the capacity region of the MAC without feedback was determined in \textcolor{black}{\cite{Ashw1971,LiaoMAC}},  \mw{single-letter expressions for the capacity   are} only known in special cases such as the two-user Gaussian MAC with perfect feedback \cite{Ozarow1984} or a class of semi-deterministic MACs 
	\cite{willems82class} with one-sided perfect feedback. \textcolor{black}{In \cite{Kramer2003}, Kramer derived a multi-letter characterization of the capacity region of a general MAC with  perfect feedback. 
		For most channels it seems however challenging to evaluate this multi-letter characterization even numerically.  
		In contrast, various inner and outer bounds on the capacity region of the  MAC with generalized or perfect feedback are known. Outer bounds are typically based on the dependence balance bound idea by Hekstra and Willems \cite{Hekstra89}, see also \cite{Sula}. 
		Various inner bounds were proposed based on schemes  that each Tx decodes part of the data sent by the other Tx, which allows the two Txs to cooperatively resend these data parts in the next block using a more efficient coding scheme, see  \cite{CoverLeung, willems1983achievable, GW,Carleial, Hekstra89}.  The one most relevant  to our work is Willems's inner bound  \cite{willems1983achievable}, which we explain in more detail in the following subsection. }
	
	\subsection{Willems' Coding Scheme with Generalized Feedback and the ISAC extension}

	\begin{figure}[h!]
		\includegraphics[scale=0.7]{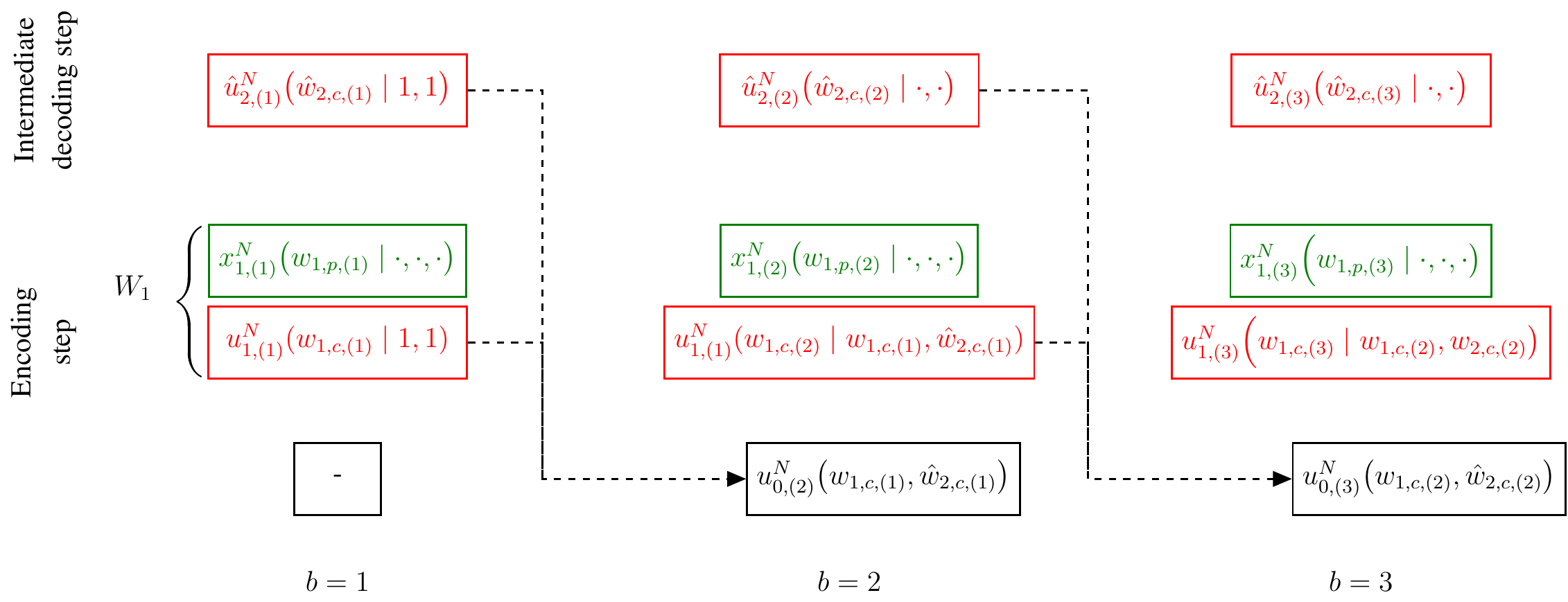}
		\caption{Operations at Tx~1 in Willems' scheme during the first three blocks. After each block $b$ Tx 1 decodes message $W_{2,c,(b)}$ based on its generalized feedback output $Z_{1,(b)}^N$. The decoded message is then retransmitted in block $b+1$ jointly with $W_{1,c,(b)}$.}
		\label{Fig:Willem's}
	\end{figure}

	Willems' scheme splits the blocklength $n$ into $B+1$ blocks of length $N=n/(B+1)$ each.  Accordingly, throughout, we let  $X_{1,(b)}^N, X_{2,(b)}^N, S_{1,(b)}^N, S_{2,(b)}^N,  Z_{1,(b)}^N, Z_{2,(b)}^N, Y_{(b)}^N$ denote the block-$b$ inputs, states and outputs, e.g., $S_{1,(b)}^N:=(S_{1(b-1)N+1}, \ldots, S_{1, bN})$. We also represent the two messages $W_1$ and $W_2$ in a one-to-one way as the $2B$-length tuples 
	\begin{equation}
		W_{k}=(W_{k,c,(1)},\ldots, W_{k,c,(B)}, W_{k,p,(1)}, \ldots, W_{k,p,(B)}), \qquad k\in\{1,2\}, 
	\end{equation}
	where all pairs  $(W_{k,c,(b)}, W_{k,p,(b)})$ are independent and uniformly distributed over $\left[2^{N\bar {R}_{k,c}}\right]\times \left[2^{N\bar {R}_{k,p}}\right]$ for $\bar{R}_{k,c}\triangleq \frac{B+1}{B}R_{k,c}$ and  $\bar{R}_{k,p}\triangleq \frac{B+1}{B}R_{k,p}$ and $R_{k,c}+R_{k,p}=R_k$.
	
	An independent superposition code is constructed for each block $b$ (see also Figure~\ref{Fig:Willem's}): 
	\begin{itemize}
		\item A lowest-level code $\mathcal{C}_{0,(b)}$ consisting of $2^{N\bar {R}_{1,c} } \cdot 2^{N\bar {R}_{2,c}}$ codewords $u_{0,(b)}(w_{1,c}, w_{2,c})$  is constructed by drawing all entries i.i.d. according to a auxiliary pmf $P_{U_0}$.

		\item 	
		\textcolor{black}{At the lowest level of encoding, we apply superposition coding to combine two codebooks $\{u^N_{k,(b)}(w'_{k,c}\mid w_{1,c},w_{2,c})\}$ onto each codeword $u_{0,(b)}^N(w_{1,c}, w_{2,c})$,}
		for $k\in\{1,2\}$ and  $w'_{k,c}\in[2^{NR_{k,c}}]$, by drawing the $i$-th entry  of each codeword according to $P_{U_k\mid U_0}(\cdot \mid u_0)$ where  $u_0$ denotes the $i$-th entry of  $u_0^N(w_{1,c},w_{2,c})$.
		\item
		\textcolor{black}{For each second-layer codeword $u^N_{k,(b)}(w'_{k,c}| w_{1,c},w_{2,c})$, we apply superposition coding by drawing the $i$-th entry of a codebook ${x_{k,(b)}^N(w_{k,p}'| w_{k,c}', w_{1,c}, w_{2,c})}$ according to $P_{X_k\mid U_0U_k}(\cdot \mid u_0,u_k)$, where $k\in{1,2}$ and $w'_{k,p}\in[2^{NR_{k,p}}]$ and $u_k$ represents the $i$-th entry of $u^N_{k,(b)}(w'_{k,c}\mid w_{1,c},w_{2,c})$.}
	\end{itemize}  
	
	As depicted in Figure~\ref{Fig:Willem's}, in Willems' scheme, Tx~$1$ sends the following block-$b$ channel inputs 
	\begin{equation}
		x_{1,(b)}^N=x_{1,(b)}^N\left( W_{1,p,(b)} \Big| W_{1,c,(b)}, W_{1,c,(b-1)}, \hat{W}_{2,c,(b-1)} \right),\qquad b\in\{1,\ldots, B+1\},
	\end{equation}
	where $ \hat{W}_{2,c,(b-1)}$ denotes the message part that Tx~1 decodes after reception of the block-$(b-1)$ generalized feedback signal $Z_{1,(b-1)}^N$, e.g., through a joint typicality decoding rule. Also, we set  throughout ${W}_{k,c,(0)}= \hat{W}_{k,c,(0)}=W_{k,p,(B+1)}=1$, for $k\in\{1,2\}$.
	
\begin{figure}[h!]
		\centering
		\includegraphics[scale=0.8]{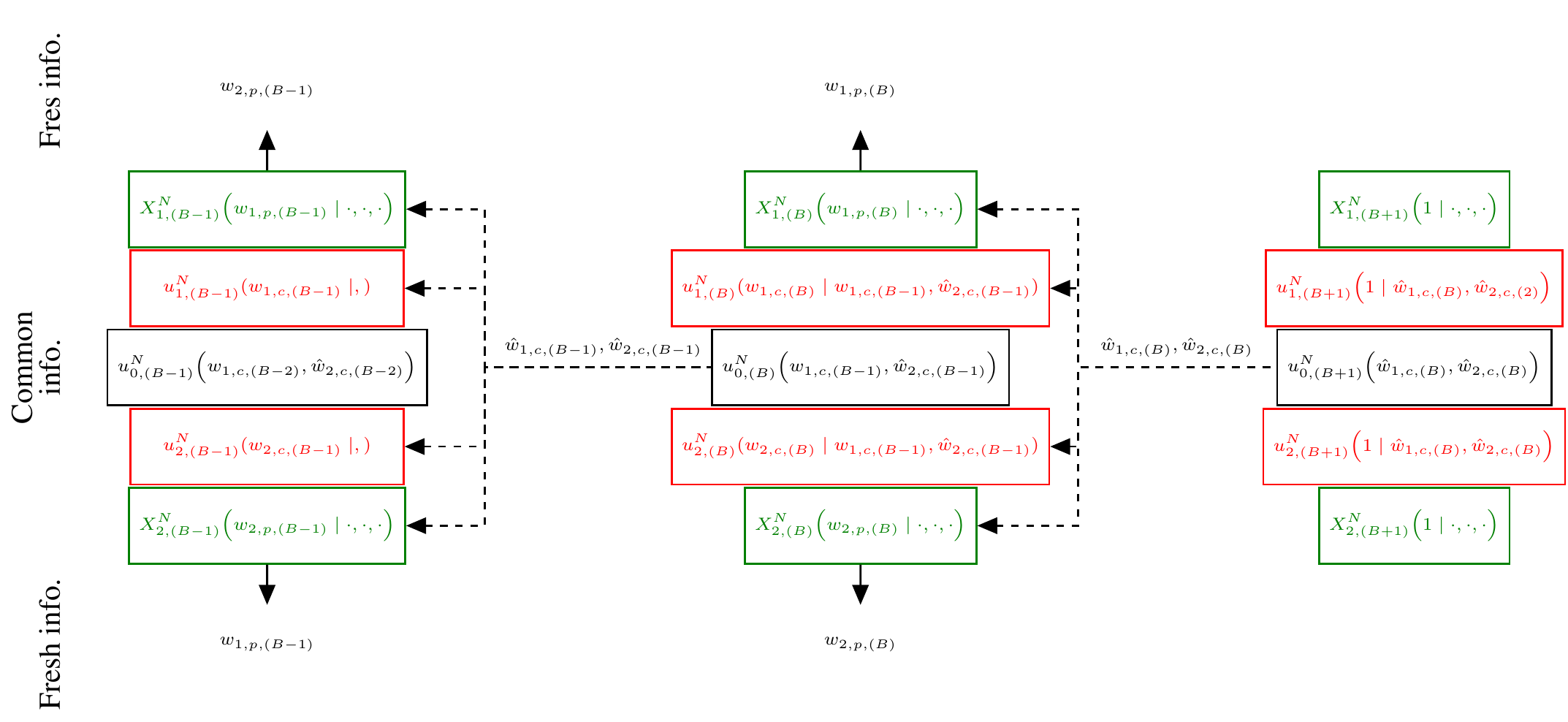}
		\caption{Backward decoding procedure at the Rx in Willems' scheme.  The pair of common messages $(W_{1,c,(b-1)},W_{2,c,(b-1)})$ and private messages $(W_{1,p,(b)},W_{2,p,(b)})$  are jointly decoded based on the block-$b$ outputs $Y_{(b)}^N$ and using the previously decoded  $(\hat W_{1,c,(b)},\hat W_{2,c,(b)})$.}
		\label{fig:decoding}
	\end{figure}

	Decoding at the Rx is performed backwards, starting with the last block $B+1$ based on which the Rx  decodes  the pair of common messages $(W_{1,c,(B)}, W_{2,c,(B)})$ using for example a joint-typicality decoder. It then uses knowledge of these common messages and the outputs in block $B$ to decode the block-$B$ private messages $(W_{1,p,(B)}, W_{2,p,(B)})$ and the block $(B-1)$ common messages $(W_{1,c,(B-1)}, W_{2,c,(B-1)})$, etc. 
	The backward decoding procedure  is also depicted in Figure~\ref{fig:decoding}. 
	
	As Willems showed, his scheme can achieve the following rate-region. 
	\begin{theorem}[Willems' Achievable Region \cite{willems1983achievable}]\label{thm:Willems}
		Any nonnegative rate-pair $(R_1,R_2)$ is achievable over the MAC with generalized feedback if it  satisfies the following inequalities
		\begin{IEEEeqnarray}{rCl}
			R_k&\leq & I(X_k; Y\mid X_{\bar{k}}U_kU_0)+I(U_k;Z_{\bar{k}}\mid X_{\bar{k}}U_0), \qquad k\in\{1,2\},\label{eq:1}
			\\
			R_1+R_2&\leq& I(X_1X_2;Y),
			\\
			R_1+R_2&\leq& I(X_1X_2;Y\mid U_0U_1U_2)
			+I(U_1;Z_2\mid X_2U_0)
			+I(U_2;Z_1\mid X_1U_0),\label{Willem}
		\end{IEEEeqnarray}	
		for some choice of  pmfs $P_{U_0}, P_{U_1\mid U_0} ,
		P_{U_2\mid U_0},P_{X_1\mid U_0 U_1},
		P_{X_2\mid U_0 U_2},$  and where above mutual informations are calculated according to the  pmf $P_{U_0}P_{U_1\mid U_0} 
		P_{U_2\mid U_0}P_{X_1\mid U_0 U_1}
		P_{X_2\mid U_0 U_2}
		P_{S_1S_2}$
		$P_{YZ_1Z_2\mid S_1S_2X_1X_2}$.  \textcolor{black}{One hereby can restrict to auxiliary variables over alphabets of sizes $|\Uc_k| \leq ( |\Xc_k|+1)|\Uc_0|$, for $k=1,2$, and $|\Uc_0| \leq   |\Xc_1||\Xc_2|+1$.} 
		%
	\end{theorem}
	
	Kobayashi et al. \cite{kobayashi2019joint} extended Willems' scheme to a ISAC scenario by adding a state estimator at the two Txs. Specifically, for any block $b$ each Tx $k$ applies the symbol-per-symbol estimation 
	\begin{equation}
		\hat{s}_{k,(b)}^N= \tilde{\phi}_{k}^{*\otimes N} \left( x_{k,(b)}^N, z_{k,(b)}^N, u_{\bar{k},(b)}^N\left( W_{\bar{k},c,(b)}\  \Big|\  W_{k,c,(b-1)}, \hat{W}_{\bar{k},c,(b-1)}\right)\right), \qquad b\in\{1,\ldots, B\},
	\end{equation}
	where $\tilde{\phi}_{k}^{*}$ denotes the optimal estimator  of $S_{k}$ based on the tuple $(X_k,Z_k,U_{\bar{k}})$:
	\begin{IEEEeqnarray}{rCl}\label{Th:distortion:MAC:Mari}
		\tilde{\phi}_{k}^*(x_k,z_k,u_{\bar{k}}):= 
		\textnormal{arg}\min_{s_k'\in \hat{\mathcal{S}_k}} \sum_{s_k\in \mathcal{S}_k}  P_{S_k|X_kZ_kU_{\bar{k}}}(s_k|x_k,z_k,u_{\bar{k}})\;  d_k(s_k,  s_k').
	\end{IEEEeqnarray}
	Thus,  any of the two Txs bases its state-estimation not only on its inputs and outputs  of a given block but also on the codeword that it decoded from the other Tx. 
	
	For the last block $B+1$, Tx $k$ can produce any trivial estimate, e.g.,  $\hat{s}_{k,(B+1)}^N$ because its influence on the average distortion vanishes as the number of blocks  grows, $B\to \infty$.
	
	Combining  the described state-estimation with Willems' scheme, the following rate-distortion region can be shown to be achievable. 

	\begin{theorem}\label{cor:MAC:Mari}[Kobayashi et al.'s ISAC region \cite{kobayashi2019joint}]	A rate-distortion tuple $(R_1, R_2, D_1, D_2)$ is achievable if it
		satisfies \textcolor{black}{\eqref{eq:1}--\eqref{Willem}}
		and
		\begin{eqnarray}\label{eq:distcor:MAC:Mari}
			\textnormal{E}\Big[d_k\left(S_k, \textcolor{black}{\tilde{\phi}_k^*\left(X_k, Z_k,U_{\bar{k}}\right)}\right)\Big]\leq D_k, \quad k=1,2,
		\end{eqnarray}
		for some choice of pmfs $P_{U_0},P_{U_1\mid U_0},P_{U_2\mid U_0},P_{X_1\mid U_1U_0},P_{X_2\mid U_2U_0}$.
		
	\end{theorem}

	\subsection{Our Proposed Collaborative ISAC Scheme}

	We present our collaborative ISAC scheme. It extends the scheme in \cite{kobayashi2019joint} in that the second-layer codeword of Willems' code construction is not only used to transmit data but also compression information useful for state sensing. 
	\textcolor{black}{Each Tx generates compression information, which is primarily intended to be used by the other Tx to improve its sensing performance.}
	In our scheme, the Rx however also decodes this information and uses it to improve its decoding performance. 

	\begin{figure}[t]
		\centering
		\includegraphics[scale=0.8]{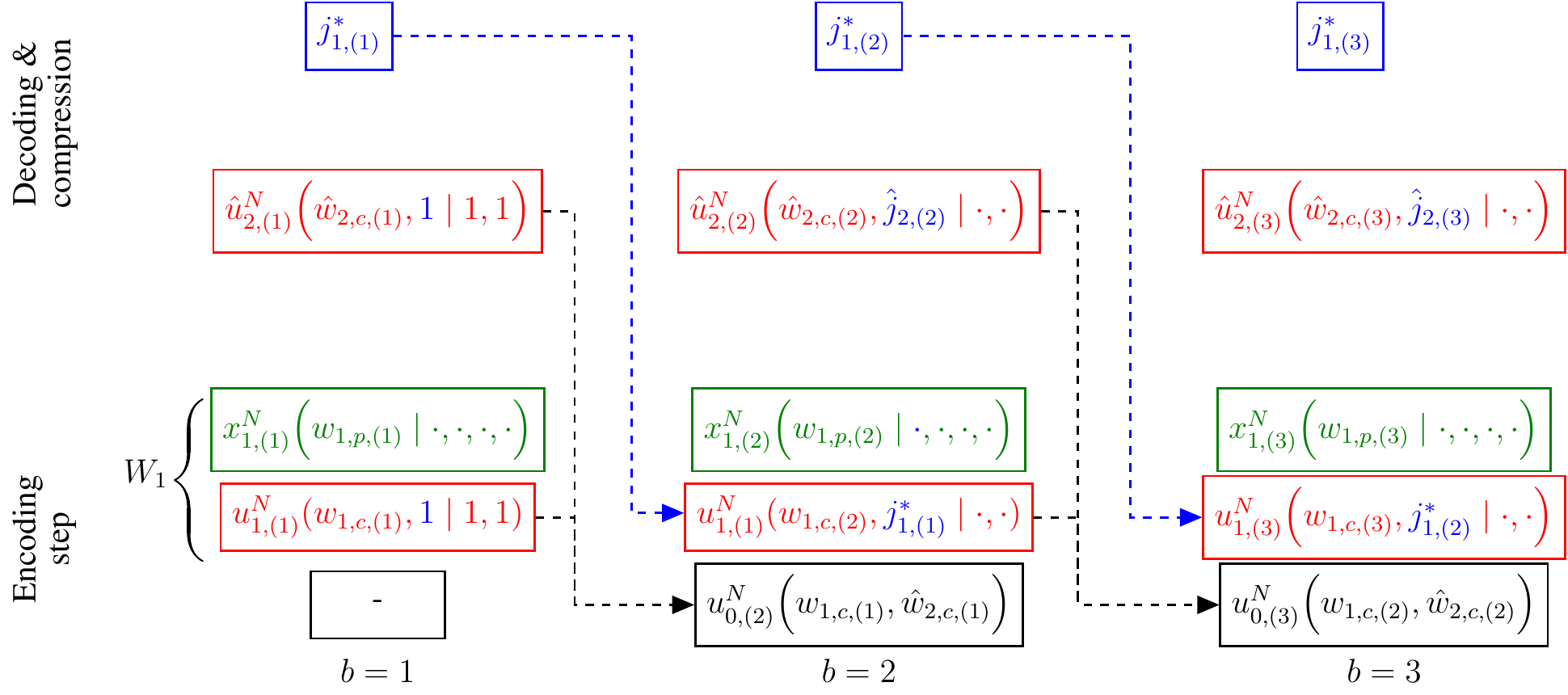}
		\caption{Our proposed  scheme at Tx~1 during the firts three blocks}
		\label{Fig:MACsensing}
	\end{figure}
	\subsubsection{Code construction} Choose pmfs $P_{U_0},P_{U_1\mid U_0}
	P_{U_2\mid U_0},
	P_{X_1\mid U_1U_0},P_{X_2\mid U_2U_0}$, and define the  pmf 
	\begin{IEEEeqnarray}{rCl}\label{eq:MACpmf}
		P_{U_0U_1U_2X_1X_2S_1S_2YZ_1Z_2V_1V_2}&=&P_{U_0}P_{U_1\mid U_0}
		P_{U_2\mid U_0}
		P_{X_1\mid U_1U_0}P_{X_2\mid U_2U_0}
		P_{S_1S_2}
		P_{YZ_1Z_2\mid X_1X_2S_1S_2}
		\nonumber\\
		&&\hspace{5cm}P_{V_1|X_1Z_1U_2U_0}\textcolor{black}{P_{V_2|X_2Z_2U_1U_0}}.
	\end{IEEEeqnarray}
	Employ Willems' three-level superposition code construction for the given choice of  pmfs, 
	except that each second-layer codeword is indexed by a pair of indices. We thus denote the second-layer codewords by $u^N_{k,(b)}(w'_{1,c}, j_{1} \mid w_{1,c},w_{2,c})$ and  $u^N_{2,(b)}(w'_{2,c}, j_{2} \mid w_{1,c},w_{2,c})$ and accordingly the corresponding third-layer codewords by $x_{1,(b)}^N(w_{1,p}'| w_{1,c}', j_1, w_{1,c}, w_{2,c})$ and 
	$x_{1,(b)}^N(w_{2,p}'| w_{2,c}', j_2, w_{1,c}, w_{2,c})$, where the indices $j_1$ and $j_2$  take value in the sets $[2^{nR_1'}]$ and $[2^{nR_2'}]$ for some positive auxiliary rates $R_{1,v}$ and $R_{2,v}$. 
	
	
	We further construct a compression codebook for each block and each  of the two Txs, For each $b\in\{1,\ldots, B\}$ and each sixtuple $(w_{1,c}, w_{2,c},  w'_{1,c}, j_1 w'_{2,c}, j_2)\in[2^{NR_{1,c}}]\times [2^{NR_{2,c}}]\times [2^{NR_{1,c}}]\times[2^{NR_{1,v}}]\times  [2^{NR_{2,c}}]\times [2^{NR_{2,v}}]$  we generate a  sequence $v_{1,(b)}^N(j'_{1}\mid  w'_{1,c}, j_{1}, w'_{2,c}, j_2,w_{1,c}, w_{2,c})$ for each $j_1' \in [2^{NR_{1,v}}]$ and  a  sequence $v_{2,(b)}^N(j'_{2}\mid w'_{1,c}, j_{1}, w'_{2,c}, j_2, w_{1,c}, w_{2,c})$ for each $j_2' \in [2^{NR_{2,v}}]$. The sequences $v_{1,(b)}^N(j'_{1}\mid w'_{1,c}, j_{1}, w'_{2,c}, j_2,w_{1,c}, w_{2,c})$ and  $v_{2,(b)}^N(j'_{2}\mid w'_{1,c}, j_{1}, w'_{2,c}, j_2,w_{1,c}, w_{2,c})$  are obtained by drawing  their $i$-th entries  according to $P_{V_1\mid U_0U_1 U_2}(\cdot \mid  u_0,u_1 ,u_2)$   and  $P_{V_2\mid U_0U_1 U_2}(\cdot \mid  u_0,u_1, u_2)$, respectively, for  $u_0, u_1,u_2$ denoting the $i$-th entries of the sequences  $u^N_{0,(b)}(w_{1,c},w_{2,c})$,   $u^N_{1,(b)}(w'_{1,c}, j_{1}\mid w_{1,c},w_{2,c})$, and $u^N_{2,(b)}(w'_{2,c}, j_{2}\mid w_{1,c},w_{2,c})$.
	\medskip
	
	\subsubsection{Operations at the Txs}
	In each block $b$, Tx~$k$  sends the block-$b$ sequence
	\begin{equation}
		X_{k,(b)}^N=x_{k,(b)}^N\left( W_{k,p,(b)} \, \Big| \, W_{k,c,(b)}, J_{k,(b-1)}^*, W_{k,c,(b-1)}, \hat{W}^{(k)}_{\bar{k},c,(b-1)} \right),
	\end{equation}
	where Tx~$k$ generates the indices $J_{k,(b-1)}^*$ and $\hat{W}_{\bar{k},c,(b-1)}$  during a joint decoding and compression step at the end of block $b-1$ as follows.  (For convenience we again set $W_{k,p,(B+1)}=W_{k,c,(0)}=\hat{W}_{\bar{k},c,(0)}^{{k}}=J_{k,(B+1)}^{\bar{k}}=1$.) 
	
	After receiving the generalized feedback signal $Z_{k,(b-1)}^N$, Tx~$k$ looks for a triple of indices $j_k^*$,  $\hat{w}_{\bar{k}}$, and $\hat{j}_{\bar{k}}$ satisfying the   joint typicality check \eqref{typ1:enc_1}, 
	
	\begin{figure*}[t]
		\begin{IEEEeqnarray}{rCl}\label{typ1:enc_1}
			\lefteqn{\Big(
				u_{0,(b-1)}^N\Big(W_{1,c,(b-2)}, \hat{W}_{2,c,(b-2)}^{(1)}\Big),
				\;
				u^N_{1,(b-1)}\Big(W_{1,c,(b-1)}, J^*_{1,(b-2)}\; \Big| \;
				W_{1,c,(b-2)}, \hat{W}_{2,c,(b-2)}^{(1)}
				\Big)} \quad 
			\nonumber
			\\
			&&
			u_{2,(b-1)}^N\Big( \hat{w}_{2},\hat{j}_{2} \;
			\Big| \;W_{1,c,(b-2)}, \hat{W}_{2,c,(b-2)}^{(1)}
			\Big), \; 
			\nonumber
			\\
			&&
			x_{1,(b-1)}^N\Big(W_{1,p,(b-1)}\;\Big| \;
			W_{1,c,(b-1)}, J^*_{1,(b-2)}, W_{1,c,(b-2)}, \hat{W}_{2,c,(b-2)}^{(1)}\Big), 
			\nonumber
			\\
			&& v_{1,(b-1)}^N\Big(j^*_{1}\;\Big|\; {J^*_{1,(b-2)}}, 
			W_{1,c,(b-1)}, \hat{w}_{2}, \hat{j}_{2} ,
			W_{1,c,(b-2)}, \hat{W}_{2,c,(b-2)}^{(1)}
			\Big), Z^N_{1,(b-1)}
			\Big) \in \mathcal{T}^N_{\epsilon}(P_{U_0U_1U_2X_1 V_1 Z_1}) \IEEEeqnarraynumspace
		\end{IEEEeqnarray}
		\hrule
		\vspace{-3mm}
	\end{figure*} 
	and if $b > 2$ also the typicality check~\eqref{typ2:enc_1},  
	\begin{figure*}[t]
		
		\begin{IEEEeqnarray}{rCl}\label{typ2:enc_1}
			\lefteqn{ \Big(u_{0,(b-2)}^N\Big(
				W_{1,c,(b-3)}, 
				\hat{W}_{2,c,(b-3)}^{(1)}\Big), \; 
				u_{1,(b-2)}^N\Big(
				W_{1,c,(b-2)},
				{J}_{1,(b-2)}^*
				\; \Big| \; 
				W_{1,c,(b-3)}, 
				\hat{W}_{2,c,(b-3)}^{(1)}
				\Big) ,}
			\nonumber	\quad	\\
			&&u_{2,(b-2)}^N\Big( \hat{W}_{2,c,(b-2)}^{(1)}, \hat{J}_{2,(b-3)}^{(1)}\; \Big| \; W_{1,c,(b-3)}, \hat{W}_{2,c,(b-3)}^{(1)}
			\Big), \;
			\nonumber
			\\
			&& 
			x_{1,(b-2)}^N\Big(W_{1,p,(b-2)}\; \Big| \; W_{1,c,(b-2)}, J^*_{1,(b-3)}, W_{1,c,(b-3)}, \hat{W}_{2,c,(b-3)}^{(1)}\Big), 
			\nonumber		\\
			&&  v_{2,(b-2)}^N\Big(
			\hat{j}_{2} 
			\; \Big| \; 
			W_{1,c,(b-2)}, 
			J^*_{1,(b-3)}
			,
			\hat{W}_{2,c,(b-2)}^{(1)},\hat{J}_{2,(b-3)}^{(1)},W_{1,c,(b-3)}, \hat{W}_{2,c,(b-3)}^{(1)}\Big), \;  
			\nonumber
			\\
			&&\hspace{7cm}
			Z^N_{1,(b-2)} \Big)
			\in \mathcal{T}^N_{\epsilon}(P_{U_0U_1U_2X_1 V_2 Z_1}).
		\end{IEEEeqnarray}
		\hrule
	\end{figure*}
	which are displayed on  top of the page. It randomly picks one of these triples  and sets 
	\begin{equation}
		J_{1,(b-1)}^*=j_1^*, \qquad \hat{W}_{2,(b-1)}^{(1)}=\hat{w}_2,\qquad \hat{J}_{2,(b-2)}^{(1)}=\hat{j}_2.
	\end{equation} 
	
	
	Tx~$k$  also produces  the block-$b$ state estimate 
	\begin{IEEEeqnarray}{rCl}
		\hat{s}_{k,(b)}^n&=&
		\phi_k^{*\otimes N}
		\Big
		( x_{k,(b)}^N\left( W_{k,p,(b)}| W_{k,c,(b)}, J_{k,(b-1)}, W_{k,c,(b-1)}, \hat{W}^{(k)}_{\bar{k},c,(b-1)} \right), \nonumber \\
		&&\hspace{1.5cm}
		z_{k,(b)}^N,u_{\bar{k},(b)}^N\left( W_{k,c,(b)} \mid J_{k,(b-1)}, W_{k,c,(b-1)}, \hat{W}^{(k)}_{\bar{k},c,(b-1)} \right),  
		\nonumber \\
		&&\hspace{2cm}
		v_{\bar{k},(b)}^N\left( J_{k,(b)} \mid W_{k,c,(b)}, J_{k,(b-1)}, W_{k,c,(b-1)}, \hat{W}^{(k)}_{\bar{k},c,(b-1)} \right)
		\Big)
	\end{IEEEeqnarray}
	where 
	\begin{IEEEeqnarray}{rCl}\label{eq:opt_estimator}
		{\phi}_{k}^*(x_k,z_k,u_{\bar{k}}, v_{\bar{k}}):= 
		\textnormal{arg}\min_{s_k'\in \hat{\mathcal{S}_k}} \sum_{s_k\in \mathcal{S}_k}  P_{S_k|X_kZ_kU_{\bar{k}}V_{\bar{k}}}(s_k|x_k,z_k,u_{\bar{k}}, v_{\bar{k}})\;  d_k(s_k,  s_k').
	\end{IEEEeqnarray}
	Without loss in performance as $B\to\infty$, the estimate in the last block $B+1$ can again be set  to a dummy sequence. 

	\medskip
	
	\subsubsection{Decoding at the Rx}
	Decoding at the Rx is similar to Willems' scheme and uses backward decoding. The difference is that   the Rx in block $b$ not only decodes the message tuple  $(W_{1,p,(b)}, W_{2,p,(b)},W_{1,c,(b-1)}, W_{2,c,(b-1)})$ but also the compression indices $J_{1,(b-1)}^*$ and $J_{2,(b-2)}^*$. 
	Specifically, in a generic block $b\in\{2,\ldots,B\}$, 
	%
	%
	the Rx looks for a unique sixtuple $(w_{1,p},w_{2,p},w_{1,c}, w_{2,c}, j_{1}, j_2) 
	\in [2^{NR_{1,p}}]
	\times
	[2^{NR_{2,p}}]\times 
	[2^{NR_{1,c}}] \times [2^{NR_{2,c}}] \times 
	[2^{NR_{1,v}}]
	\times
	[2^{NR_{2,v}}]
	$  satisfying
	\begin{IEEEeqnarray}{rCl}\label{typ:dec_b}
		&&\hspace{-0.5cm}	\Bigg(
		u^N_{0,b}(w_{1,c}, w_{2,c}),\;
		u^N_{1,(b)}\Big(\hat{W}_{1,c,(b)},j_{1} \; \Big | \;  w_{1,c}, w_{2,c} \Big),\;
		u^N_{2,(b)}\Big(\hat{W}_{2,c,(b)}, j_{2} \; \Big | \;  w_{1,c}, w_{2,c}\Big),
		\nonumber \\
		& & \quad x^N_{1,(b)}\Big(w_{1,p} \; \Big | \;  \hat{W}_{1,c,(b)}, j_{1}, w_{1,c}, w_{2,c}\Big), \; 
		x^N_{2,(b)}\Big( w_{2,p} \; \Big | \;  \hat{W}_{2,c,(b)}, j_{2}, w_{1,c}, w_{2,c}\Big),
		\nonumber		\\
		&&\quad\hspace{2cm}
		v_{1,(b)}^N\left(\hat{J}_{1,(b)} \; \Big | \;  
		\hat{W}_{1,c,(b)}, j_1,  \hat{W}_{2,c,(b)}, j_2,
		w_{1,c}, w_{2,c}
		\right), 
		\nonumber		\\
		&&\quad\hspace{2cm}
		v^N_{2,(b)}\left(\hat{J}_{2,(b)} \; \Big | \;  \hat{W}_{1,c,(b)}, j_1,  \hat{W}_{2,c,(b)}, j_2,w_{1,c}, w_{2,c} 
		\right), \;
		Y^N_{(b)} \Bigg) { \in \Tc^{N}_{2\epsilon}(P_{U_0U_1U_2X_1X_2 Y}).}\;\;
	\end{IEEEeqnarray}
	If such a  unique sixtuple exists, it sets $\hat{W}_{1,c,(b-1)}=w_{1,c}$, $\hat{W}_{1,p,(b)}=w_{1,p}$, $\hat{W}_{2,c,(b-1)}=w_{2,c}$, $\hat{W}_{2,p,(b)}=w_{2,p}$, $\hat{J}_{1,(b-1)}=j_1$, and $\hat{J}_{2,(b-1)}=j_2$. Otherwise it declares  an error.
	
	
	%
	The Rx finally declares the messages $\hat{W}_1$ and $\hat{W}_2$ that correspond to the produced guesses $\{(\hat{W}_{k,p,(b)}, \hat{W}_{k,c,(b)})\}$.
	%

	In \lo{Appendix~\ref{app:analysis}} \sh{\cite{ArxivMacIt}} we show that as $N\to \infty$ and $B\to \infty$, the described scheme achieves vanishing probabilities of error, the compressions are successful with probability $1$,  and the asymptotic expected distortions are bounded by $D_1$ and $D_2$ whenever $B$ is sufficiently large and 
	\begin{subequations}\label{eq:subrates}
		\begin{IEEEeqnarray}{rCl}\label{beforeFME}
			R_{k,v} &>&  I(V_k;X_kZ_k \mid \underline{U}) \label{eq:Rkv}\\
			{R_{\bar{k},v}+} R_{k,c}&<& I(U_kV_{\bar{k}};X_{\bar{k}}Z_{\bar{k}}\mid U_0U_{\bar{k}})\label{eq:Rkc}\\
			%
			{R_{1,v}+} R_{2,v}+R_{k,c}&<& I(U_kV_{\bar{k}}; X_{\bar{k}}Z_{\bar{k}}\mid U_0U_{\bar{k}})
			+I(V_k;X_{\bar{k}}Z_{\bar{k}}\mid   \underline{U})\label{eq:RkvRkc}
			\\
			R_{k,p} &< & I(X_k ; YV_1V_2\mid \underline{U} X_{\bar{k}} ) \label{eq:Rkp}
			\\
			R_{k,v}+R_{k,p}&<& I(X_k; Y \mid U_0 X_{\bar{k}} ) + I( V_2; X_1X_2YV_1   \mid \underline{U})+I( V_1; X_1X_2Y \mid   \underline{U}) \IEEEeqnarraynumspace
			\\
			R_{k,v}+R_{k,p}+R_{\bar{k},p}&<&
			I(X_1X_2; Y \mid U_0 U_{\bar{k}}) + I( V_2 ;X_1X_2YV_1 \mid \underline{U})
			\nonumber\\
			&&\hspace{5cm}	+ I( V_1; X_1X_2Y \mid  \underline{U}) \\
			R_{1,p}+R_{2,p}& <& I(X_1 X_2; YV_1V_2 \mid \underline{U}  )\label{eq:R1pR2p}
			\\
			R_{1,v} +R_{1,p}+ R_{2,v}+R_{2,p}& <& 
			I(X_1X_2; Y \mid U_0 )  + I( V_1; X_1X_2Y \mid   \underline{U}) 	+ I( V_2; X_1X_2YV_1   \mid \underline{U}) 
			\\
			R_{1,v}+R_1 +R_{2,v}+R_2  & <&  I(X_1X_2; Y ) + I( V_1; X_1X_2Y \mid   \underline{U}) 
			+ I( V_2; X_1X_2YV_1 \mid \underline{U}), \label{eq:R1R2R2v}
			\IEEEeqnarraynumspace
		\end{IEEEeqnarray}
		where $\underline{U}\eqdef (U_0,U_1,U_2)$ and 
		\begin{eqnarray}\label{Th:distortion:MAC}
			&	 \textnormal{E}[d_k(S_k, \phi^*_{k}(X_k, Z_k, U_{\bar{k}}, V_{\bar{k}}
			)
			]\leq D_k,  \quad k=1,2,\label{eq:Dk}
		\end{eqnarray}
		for $\phi_k^*$  defined in \eqref{eq:opt_estimator}.  		
	\end{subequations}
	
	Using the Fourier-Motzkin Elimination (FME) algorithm it can be shown, see \sh{\cite{ArxivMacIt}}\lo{Appendix~\ref{App:FME:MAC}}, that such a choice of rates is possible under the rate-constraints \eqref{eq:inner:MAC}.

	\begin{theorem}\label{Th:achievability:MAC}
		The capacity-distortion region $\Cc\Dc$ includes any rate-distortion tuple $(R_1, R_2, D_1, D_2)$ that for some choice of  pmfs 
		$P_{U_0}, P_{U_1\mid U_0},
		P_{U_2\mid U_0}, P_{X_1\mid U_0U_1}, P_{X_2\mid U_0U_2},$ $P_{V_1\mid U_0U_2X_1Z_1}, P_{V_2\mid U_0U_1X_2Z_2}$ and pmf $P_{U_0U_1U_2X_1X_2S_1S_2YZ_1Z_2V_1V_2}$   \textcolor{black}{as defined in \eqref{eq:MACpmf}}, 
		satisfies  Inequalities \eqref{eq:inner:MAC} on top of the next page (where $\underline{U}:=(U_0,U_1,U_2)$)
		\begin{figure*}[t]
			
			\begin{subequations}\label{eq:inner:MAC}
				\begin{IEEEeqnarray}{rCl}
					R_k&\leq&I(U_k;X_{\bar{k}}Z_{\bar{k}}\mid U_0U_{\bar{k}})
					+
					I(V_k;X_{\bar{k}}Z_{\bar{k}}\mid \underline{U} )
					-I(V_k;X_kZ_k\mid \underline{U})+\min\{
					\nonumber\\
					&&
					I(X_k;Y\mid U_0X_{\bar{k}})
					+ I( V_k; X_1X_2Y \mid \underline{U}) 	+ I( V_{\bar{k}}; X_1X_2YV_k \mid \underline{U})
					\nonumber\\
					&&\hspace{10cm}	-I(V_k;X_kZ_k\mid \underline{U}),
					\nonumber\\
					&&\hspace{0cm} I(X_1X_2;Y\mid U_0U_k)
					+I( V_k; X_1X_2Y \mid \underline{U}) 	+ I( V_{\bar{k}}; X_1X_2YV_k\mid \underline{U})
					\nonumber\\
					&&\hspace{10cm}
					-I(V_{\bar{k}};X_{\bar{k}}Z_{\bar{k}}\mid \underline{U}), \nonumber\\
					&&
					\hspace{0cm}  I(X_1X_2;Y\mid U_0)+ I( V_k; X_1X_2Y \mid \underline{U}) 	+ I( V_{\bar{k}}; X_1X_2YV_k\mid \underline{U})
					\nonumber \\
					& & \hspace{0.3cm} 
					-I(V_k;X_kZ_k\mid \underline{U}) 
					-I(V_{\bar{k}};X_{\bar{k}}Z_{\bar{k}}\mid \underline{U})   , \ I(X_k;YV_1V_2\mid \underline{U}X_{\bar{k}})
					\} , \hspace{0.1cm}  k=1,2,	\IEEEeqnarraynumspace
					\vspace{0.5cm}\\
					%
					%
					R_1+R_2&\leq& I(U_2;X_1Z_1\mid U_0U_1)
					+I(V_2;X_1Z_1\mid \underline{U})
					- I(V_2;X_2Z_2\mid \underline{U})
					\nonumber\\
					&& +I(U_1;X_2Z_2\mid U_0U_2)	+I(V_1;X_2Z_2\mid \underline{U})
					- I(V_1;X_1Z_1\mid \underline{U})+	\min\{
					\nonumber\\
					&&
					I(X_1X_2;Y \mid U_0U_2)
					+I( V_1; X_1X_2Y \mid \underline{U}) 	+ I( V_2; X_1X_2YV_1 \mid \underline{U})
					-I(V_1;X_1Z_1\mid \underline{U}) ,
					\nonumber\\
					&&
					\hspace{0cm}
					I(X_1X_2;Y\mid U_0U_1)
					+I( V_1; X_1X_2Y \mid \underline{U}) 	+ I( V_2; X_1X_2YV_1 \mid \underline{U})	
					-I(V_2;X_2Z_2\mid \underline{U}),
					\nonumber\\
					&&
					\hspace{0cm} I(X_1X_2;Y\mid U_0)
					+I( V_1; X_1X_2Y \mid \underline{U}) 	+ I( V_2; X_1X_2YV_1 \mid \underline{U})	
					\nonumber\\
					&&\hspace{8cm}	-I(V_1;X_1Z_1\mid \underline{U}) -I(V_2;X_2Z_2\mid \underline{U}),
					\nonumber\\
					&& \hspace{0cm} I(X_1X_2;YV_1V_2\mid \underline{U})
					\} \vspace{0.3cm}	\\
					R_{1}+R_{2}&\leq&
					I(X_1X_2; Y )
					+ I( V_1; X_1X_2Y \mid \underline{U}) 		-I(V_1;X_1Z_1\mid \underline{U})
					\nonumber\\
					&& \hspace{4cm}
					+ I( V_2; X_1X_2YV_1 \mid \underline{U})
					-I(V_2;X_2Z_2\mid \underline{U})\label{eq:R122}
					\nonumber\\&&
					%
				\end{IEEEeqnarray}
				and for $k=1,2$
				\begin{IEEEeqnarray}{rCl}\label{conditioncompression}
					I(U_k;X_{\bar{k}}Z_{\bar{k}}\mid U_0U_{\bar{k}})+I(V_k;X_{\bar{k}}Z_{\bar{k}}\mid \underline{U}) &\geq& I(V_k;X_kZ_k\mid \underline{U})\hspace{0cm},\\
					I(X_1X_2;Y\mid U_0)+
					I( V_1; X_1X_2Y \mid \underline{U}) 	+ I( V_2; X_1X_2YV_1 \mid \underline{U})
					& \geq& I(V_1;X_1Z_1\mid \underline{U}) 
					\nonumber\\&&\hspace{0.2cm}+I(V_2;X_2Z_2\mid \underline{U})\\
					I(X_k;Y\mid U_0X_{\bar{k}})
					+
					I( V_1; X_1X_2Y \mid \underline{U}) 	+ I( V_2; X_1X_2YV_1 \mid \underline{U}) &\geq& I(V_k;X_kZ_k\mid \underline{U}) \hspace{0.0cm}.
				\end{IEEEeqnarray}
			\end{subequations}
			\vspace{-3mm}
			\hrule
			\vspace{-2mm}
		\end{figure*}
		as well as the distortion constraints~\eqref{Th:distortion:MAC}. \textcolor{black}{It suffices to consider auxiliary random variables with alphabets of  sizes $|\Uc_0| \leq |\mathcal{X}_1||\mathcal{X}_2|+9$   and for $k=1,2$: $|\Uc_k| \leq (|\mathcal{X}_k| +9)|\Uc_0|$ and $|\mathcal{V}_k| \leq  (|\mathcal{X}_k| |\mathcal{Z}_k| |\mathcal{U}_{\bar{k}}| |\mathcal{U}_{0}| +9)$.}
	\end{theorem}
	
	Notice that Theorem~\ref{Th:achievability:MAC} recovers the previous achievable region in Theorem~\ref{cor:MAC:Mari} through the choice $V_1=V_2=$constants, which removes the collaborative sensing  between the two Txs. 

	%
	%
	%
	
	\begin{remark}[Wyner-Ziv Coding]
		In our scheme, no binning as in Wyner-Ziv coding is used for the compression of the $V_1$- and $V_2$-codewords. Instead, decoder side-information is taken into account through the additional typicality check \eqref{typ2:enc_1} and by including the $V_1$- and $V_2$-codewords in the typicality check \eqref{typ:dec_b}. These strategies are  known as implicit binning  and allow multiple decoders to exploit different levels of side-information, see  \cite{Tuncel}.
	\end{remark}
	
	\subsection{Examples}
	The following two examples show the improvement of   Theorem~\ref{Th:achievability:MAC}  over Theorem~\ref{cor:MAC:Mari}.
	\textcolor{black}{In the first example,  one of the transmitting nodes directly receives state information through its feedback, and  allows to easily illustrate the concept of collaborative sensing. The second example presents a more realistic model, and provides a more practical implementation of our collaborative sensing scheme.}
	\begin{example}
		Consider a MAC with binary input, output, and state alphabets $\Xc_1=\Xc_2=\Yc=\Sc_2=\{0, 1\}$. State $S_2\sim Ber(p_s)$, while $S_1=0$ is a constant. The channel input-output relation is described by 
		\begin{equation}\label{ex1:channel}
			Y= S_2 X_2,  
			\qquad
			(Z_1,Z_2)=(S_2,X_1).
		\end{equation} 	
		\mw{For this channel,} the following tuple
		\begin{equation}\label{eq:t1}
			(\R_1,  \R_2,  \D_1,  \D_2)= (0,  0,  0, 0),
		\end{equation}
		lies in the achievable region of Theorem~\ref{Th:achievability:MAC} through the choice $V_1=Z_1=S_2$ and $(\hat{S}_2=V_1, \hat{S}_1=0)$. Distortion $\D_2=0$ is however not achievable in Theorem~\ref{cor:MAC:Mari} because  $S_2$ is independent  of $(U_1, U_2, U_0, X_1, X_2)$ and thus of $(X_2, U_1, Z_2)$, and the \mw{optimal estimator is the trivial estimator $\hat{S}_2=\psi_2^*( X_2,Z_2,U_1)=\mathbbm{1}\{ p_s > 1/2)$ which achieves} distortion $\D_2=\min\{1-p_s, p_s\}$.
		%
	\end{example}
	
	
	\begin{example}\label{Ex:MAC:NoisyFB}
		Consider binary noise, states and channel inputs  $B_0,B_k, S_k , X_k \in \{0, 1\}$.
		The noise to the receiver $B_0$  is  Bernoulli-$t_0$, and $B_k$, the noise on the feedback to Tx $k$, is Bernoulli-$t_{k}$. All noises are independent and also 
		independent of the states    $S_1,S_2$, which are i.i.d. Bernoulli-$p_s$. 
		We can then des are described the channel as 
		\begin{IEEEeqnarray}{rCl}
			Y'&=&S_1X_1+ S_2X_2+B_0, \hspace{1cm} Y  =  (Y', S_1,S_2),\\
			Z_1  &=& S_1X_1+ S_2X_2+B_1, \hspace{1cm}
			Z_2=S_1X_1+ S_2X_2+B_2.
		\end{IEEEeqnarray}
		where \textcolor{black}{the summation operators '$+$' denote real additions.}
		In this example the Rx thus has perfect channel state-information, see also Remark~\ref{rem:state1}.
		Hamming distance is considered as a distortion measure: $d(s,\hat{s})=s\oplus\hat{s}$ where \textcolor{black}{the operator '$\oplus$' is a binary operation representing module-2 addition}.
		
		We compare Theorems \ref{cor:MAC:Mari} and \ref{Th:achievability:MAC} on the following choices of random variables. Let
		\begin{IEEEeqnarray}{rCl}
			X_k&=& \underbrace{U_0 \oplus \Sigma }_{\defeq U_1}\oplus \theta_k , \quad \textnormal{for } k\in\{1,2\}
		\end{IEEEeqnarray}
		where $U_0, \Sigma_1\Sigma_2, \theta_1,\theta_2$ are all independent Bernoulli random variables of parameters $p, q_1,q_2,r_1,r_2$.  For the evaluation of Theorem~\ref{Th:achievability:MAC} we further choose the compression random variables
		\begin{equation}\label{eq:compression} V_k=\begin{cases} \mathbbm{1}\{Z_k=1\}+2\cdot \mathbbm{1}\{Z_k=2\} & \textnormal{ if } E_k=0 \\ ``\textnormal{?}" & \textnormal{ if } E_k=1\end{cases} \qquad\forall k=\{1,2\}
		\end{equation} for a binary  $E_k$ independent of $(S_1, S_2,B_0, B_1,B_2, U_0, U_1, U_2,\Sigma_1,\Sigma_2,\theta_1,\theta_2)$.
		For this choice,  Tx $k$ conveys  information about  $Z_k$ to  Tx $\bar{k}$, which  helps this latter to better estimate its state $S_{\bar{k}}$.  For instance, when $E_1=0$, Tx-$2$ receives another noisy observation of the output which helps  it to better estimate its state, because  
		\begin{equation}
			Y= \begin{cases} 0 & \textnormal{ if } Z_2\in\{0,1\}, V_1=0 \\  1 & \textnormal{ if } V_1=1\\
				2 & \textnormal{ if } Z_2\in\{2,3\}, V_1=0\end{cases}.
		\end{equation}

	\end{example}

	\begin{figure}[h!]
		\centering
		%
		
		%
		%
		%
		%
		\definecolor{mycolor1}{rgb}{0.00000,0.44700,0.74100}%
		\definecolor{mycolor2}{rgb}{0.85000,0.32500,0.09800}%
		\begin{tikzpicture}[scale=0.8]
			
			\begin{axis}[%
				width=3.028in,
				height=2.654in,
				at={(2.511in,3.642in)},
				scale only axis,
				xmin=0.02,
				xmax=0.084,
				xlabel style={font=\color{white!15!black}},
				xlabel={$\text{D}_\text{2}$},
				ymin=0,
				ymax=1.2,
				ylabel style={font=\color{white!15!black}},
				ylabel={$\text{R}_\text{1}\text{+R}_\text{2}$},
				axis background/.style={fill=white},
				legend style={at={(0.55,0.162)}, anchor=south west, legend cell align=left, align=left, draw=white!15!black}
				]
				\addplot [color=mycolor1, line width=3.0pt]
				table[row sep=crcr]{%
					0.092	-2.22044604925031e-16\\
					0.052	-5.55111512312578e-17\\
					0.04	0\\
					0.03474	0.0105824077589774\\
					0.03474	0.0105824077589775\\
					0.03474	0.0105824077589775\\
					0.03474	0.0105824077589776\\
					0.04132	0.361236236895901\\
					0.04784	0.558204204277033\\
					0.06934	0.969128046808845\\
					0.07696	1.09633026386404\\
					0.077608	1.10171459932501\\
					0.0835	1.14863871335643\\
					0.0835	1.14863871335643\\
					0.0868	1.13753556099678\\
					0.088528	1.13009186949911\\
					0.08896	1.12749999915116\\
					0.09172	1.0766120642432\\
					0.09334	1.02914300439165\\
					0.09556	0.934157975205315\\
					0.09724	0.823921566367295\\
					0.09778	0.781536542669041\\
					0.1	0.57636898744717\\
					0.1	0.57636898744717\\
					0.1	0.576103319175841\\
					0.1	0.576103319175841\\
					0.1	0.576103319175841\\
					0.1	0.575652454603709\\
					0.1	0.575132357300687\\
					0.1	0.575132357300687\\
					0.1	0.575132357300687\\
					0.1	0.57486369047578\\
					0.1	0.572655896080304\\
					0.1	0.569483326804367\\
					0.1	0.565346974144484\\
					0.1	0.56024539693423\\
					0.1	0.55806865302353\\
					0.1	0.557012668511208\\
					0.1	0.555193306689922\\
					0.1	0.554174648006119\\
					0.1	0.552081617298862\\
					0.1	0.548289614409193\\
					0.1	0.548289614409193\\
					0.1	0.547128157087541\\
					0.1	0.547128157087541\\
					0.1	0.543292999887793\\
					0.1	0.539096564106717\\
					0.1	0.537215123867722\\
					0.1	0.530625072500171\\
					0.1	0.530625072500171\\
					0.1	0.525840840454514\\
					0.1	0.521648337402094\\
					0.1	0.521252985532249\\
					0.1	0.521252985532249\\
					0.1	0.514013722385443\\
					0.1	0.514013722385443\\
					0.1	0.508381451982377\\
					0.1	0.508381451982377\\
					0.1	0.508381451982377\\
					0.1	0.508381451982377\\
					0.1	0.50797644418321\\
					0.1	0.50797644418321\\
					0.1	0.493346042732215\\
					0.1	0.493346042732215\\
					0.1	0.485901428140046\\
					0.1	0.485901428140046\\
					0.1	0.485512557083268\\
					0.1	0.485512557083268\\
					0.1	0.485512557083268\\
					0.1	0.48517199001783\\
					0.1	0.48517199001783\\
					0.1	0.481980251949233\\
					0.1	0.481980251949233\\
					0.1	0.479583370111743\\
					0.1	0.479583370111743\\
					0.1	0.472923537189286\\
					0.1	0.472367399244587\\
					0.1	0.470099827453861\\
					0.1	0.470099827453861\\
					0.1	0.468449220044945\\
					0.1	0.464920367838359\\
					0.1	0.464920367838359\\
					0.1	0.464920367838358\\
					0.1	0.464920367838358\\
					0.1	0.457713600794981\\
					0.1	0.457713600794981\\
					0.1	0.457713600794981\\
					0.1	0.45771360079498\\
					0.1	0.456341752225864\\
					0.1	0.454685727107353\\
					0.1	0.454685727107353\\
					0.1	0.454685727107352\\
					0.1	0.452208026553418\\
					0.1	0.449420928264365\\
					0.1	0.434570321810979\\
					0.1	0.433361574151994\\
					0.1	0.425766952498131\\
					0.1	0.425766952498131\\
					0.1	0.424493620237877\\
					0.1	0.420309560395941\\
					0.1	0.419601344085339\\
					0.1	0.419601344085339\\
					0.1	0.419601344085339\\
					0.1	0.419601344085339\\
					0.1	0.416386752154004\\
					0.1	0.416386752154004\\
					0.1	0.416386752154004\\
					0.1	0.416386752154004\\
					0.1	0.413676068660746\\
					0.1	0.413676068660745\\
					0.1	0.413676068660745\\
					0.1	0.396773244377371\\
					0.1	0.396773244377371\\
					0.1	0.393799611114702\\
					0.1	0.390226314011638\\
					0.1	0.390226314011638\\
					0.1	0.384277368566017\\
					0.1	0.382675218253471\\
					0.1	0.382675218253471\\
					0.1	0.382675218253471\\
					0.1	0.376958278263917\\
					0.1	0.375650500670558\\
					0.1	0.375650500670558\\
					0.1	0.375217289721437\\
					0.1	0.375059903513028\\
					0.1	0.375059903513028\\
					0.1	0.375059903513028\\
					0.1	0.375059903513028\\
					0.1	0.369529023390588\\
					0.1	0.361735285398626\\
					0.1	0.361735285398626\\
					0.1	0.360169898238937\\
					0.1	0.358976166943764\\
					0.1	0.358976166943764\\
					0.1	0.355959007785398\\
					0.1	0.35423764807741\\
					0.1	0.348245176736094\\
					0.1	0.343148080048003\\
					0.1	0.343148080048003\\
					0.1	0.343148080048003\\
					0.1	0.33958348400881\\
					0.1	0.33958348400881\\
					0.1	0.33958348400881\\
					0.1	0.338763928120032\\
					0.1	0.336994334363306\\
					0.1	0.336994334363306\\
					0.1	0.333733054872052\\
					0.1	0.333733054872052\\
					0.1	0.333733054872052\\
					0.1	0.333733054872052\\
					0.1	0.333245604133161\\
					0.1	0.330360669873092\\
					0.1	0.330360669873092\\
					0.1	0.321691701559159\\
					0.1	0.321179089510156\\
					0.1	0.321179089510156\\
					0.1	0.320064617208594\\
					0.1	0.318423331479471\\
					0.1	0.318423331479471\\
					0.1	0.31221298490617\\
					0.1	0.31221298490617\\
					0.1	0.310645659425448\\
					0.1	0.310645659425448\\
					0.1	0.310645659425448\\
					0.1	0.307276605986977\\
					0.1	0.307276605986977\\
					0.1	0.307276605986977\\
					0.1	0.302711493656091\\
					0.1	0.302621845992337\\
					0.1	0.302621845992337\\
					0.1	0.300569577976147\\
					0.1	0.296962184875733\\
					0.1	0.279959336178251\\
					0.1	0.279959336178251\\
					0.1	0.276539070968107\\
					0.1	0.276539070968107\\
					0.1	0.276539070968106\\
					0.1	0.276407119562671\\
					0.1	0.27640711956267\\
					0.1	0.276152414357214\\
					0.1	0.268249357621367\\
					0.1	0.268249357621367\\
					0.1	0.262375227832262\\
					0.1	0.260678765618306\\
					0.1	0.249201483058166\\
					0.1	0.249201483058166\\
					0.1	0.249201483058166\\
					0.1	0.249201483058166\\
					0.1	0.243690856872702\\
					0.1	0.243690856872702\\
					0.1	0.243690856872702\\
					0.1	0.239854055147908\\
					0.1	0.239854055147908\\
					0.1	0.23439090764587\\
					0.1	0.23439090764587\\
					0.1	0.233876869250398\\
					0.1	0.233876869250398\\
					0.1	0.224395346360879\\
					0.1	0.224395346360879\\
					0.1	0.224180877688377\\
					0.1	0.213720990418992\\
					0.1	0.213720990418992\\
					0.1	0.211229299388191\\
					0.1	0.211229299388191\\
					0.1	0.211229299388191\\
					0.1	0.205274340254908\\
					0.1	0.205274340254908\\
					0.1	0.205274340254907\\
					0.1	0.205274340254907\\
					0.1	0.199748774117565\\
					0.1	0.199504380879429\\
					0.1	0.199504380879429\\
					0.1	0.192374695729069\\
					0.1	0.192374695729069\\
					0.1	0.188111927103452\\
					0.1	0.188111927103452\\
					0.1	0.185986527544492\\
					0.1	0.178767741903679\\
					0.1	0.178767741903679\\
					0.1	0.178767741903679\\
					0.1	0.178494287245091\\
					0.1	0.178494287245091\\
					0.1	0.178494287245091\\
					0.1	0.165131892508459\\
					0.1	0.165131892508459\\
					0.1	0.161347197451649\\
					0.1	0.161347197451649\\
					0.1	0.161347197451649\\
					0.1	0.161347197451649\\
					0.1	0.159643493087222\\
					0.1	0.159643493087222\\
					0.1	0.151828507846025\\
					0.1	0.150358483812268\\
					0.1	0.150358483812268\\
					0.1	0.147943660647037\\
					0.1	0.147943660647037\\
					0.1	0.147943660647037\\
					0.1	0.147792177400607\\
					0.1	0.147119954868653\\
					0.1	0.147119954868653\\
					0.1	0.147119954868653\\
					0.1	0.146306184419168\\
					0.1	0.146306184419168\\
					0.1	0.13075940413749\\
					0.1	0.13075940413749\\
					0.1	0.119538212056879\\
					0.1	0.119538212056879\\
					0.1	0.11742005464839\\
					0.1	0.11742005464839\\
					0.1	0.11742005464839\\
					0.1	0.11742005464839\\
					0.1	0.117393034048984\\
					0.1	0.117393034048984\\
					0.1	0.117393034048984\\
					0.1	0.115545088588597\\
					0.1	0.113844626934656\\
					0.1	0.113844626934656\\
					0.1	0.109597827256721\\
					0.1	0.108342271895467\\
					0.1	0.108342271895467\\
					0.1	0.101281881178936\\
					0.1	0.101281881178936\\
					0.1	0.101281881178936\\
					0.1	0.0963869157665208\\
					0.1	0.0963869157665207\\
					0.1	0.08684240745093\\
					0.1	0.0868424074509299\\
					0.1	0.081383069450145\\
					0.1	0.081383069450145\\
					0.1	0.081383069450145\\
					0.1	0.0794329310265362\\
					0.1	0.0794329310265361\\
					0.1	0.0792616693311703\\
					0.1	0.0792616693311702\\
					0.1	0.0734929118451314\\
					0.1	0.0734929118451313\\
					0.1	0.0734929118451313\\
					0.1	0.0734929118451312\\
					0.1	0.0714034771128363\\
					0.1	0.0663260599786661\\
					0.1	0.0663260599786661\\
					0.1	0.0620144273955514\\
					0.1	0.0620144273955514\\
					0.1	0.0562917808528765\\
					0.1	0.0562917808528764\\
					0.1	0.0554438074892196\\
					0.1	0.0554438074892196\\
					0.1	0.0554438074892195\\
					0.1	0.0489215119656336\\
					0.1	0.0489215119656335\\
					0.1	0.042978250073743\\
					0.1	0.042978250073743\\
					0.1	0.0393276499961932\\
					0.1	0.0393276499961931\\
					0.1	0.0332091269689512\\
					0.1	0.0295657690418726\\
					0.1	0.0295657690418725\\
					0.1	0.0295657690418724\\
					0.1	0.0276419390245821\\
					0.1	0.027641939024582\\
					0.1	0.0257411542548229\\
					0.1	0.0257411542548228\\
					0.1	0.0243098480618653\\
					0.1	0.0243098480618652\\
					0.1	0.0164599544811221\\
					0.1	0.016459954481122\\
					0.1	0.009605733799503\\
					0.1	0.00960573379950298\\
					0.1	0.00960573379950292\\
					0.1	0.00669483081631583\\
					0.1	0.00669483081631578\\
					0.1	0\\
					0.098	-8.32667268468867e-17\\
					0.092	-2.22044604925031e-16\\
				};
				\addlegendentry{Theorem \ref{Th:achievability:MAC}}
				
				\addplot [color=mycolor2, line width=2.0pt]
				table[row sep=crcr]{%
					0.1	0\\
					0.098	-8.32667268468867e-17\\
					0.092	-2.22044604925031e-16\\
					0.052	-5.55111512312578e-17\\
					0.04	0\\
					0.046	0.302711493656091\\
					0.05086	0.481710769965457\\
					0.055228	0.608401090919568\\
					0.06064	0.760047666852389\\
					0.06934	0.969128046808845\\
					0.07696	1.09633026386404\\
					0.077608	1.10171459932501\\
					0.0835	1.14863871335643\\
					0.0868	1.13753556099678\\
					0.088528	1.13009186949911\\
					0.08896	1.12749999915116\\
					0.09172	1.0766120642432\\
					0.09334	1.02914300439165\\
					0.09556	0.934157975205315\\
					0.09724	0.823921566367295\\
					0.09778	0.781536542669041\\
					0.1	0.57636898744717\\
					0.1	0.57636898744717\\
					0.1	0.576103319175841\\
					0.1	0.576103319175841\\
					0.1	0.576103319175841\\
					0.1	0.575652454603709\\
					0.1	0.575132357300687\\
					0.1	0.575132357300687\\
					0.1	0.575132357300687\\
					0.1	0.57486369047578\\
					0.1	0.572655896080304\\
					0.1	0.569483326804367\\
					0.1	0.565346974144484\\
					0.1	0.56024539693423\\
					0.1	0.55806865302353\\
					0.1	0.557012668511208\\
					0.1	0.555193306689922\\
					0.1	0.554174648006119\\
					0.1	0.552081617298862\\
					0.1	0.548289614409193\\
					0.1	0.548289614409193\\
					0.1	0.547128157087541\\
					0.1	0.547128157087541\\
					0.1	0.543292999887793\\
					0.1	0.539096564106717\\
					0.1	0.537215123867722\\
					0.1	0.530625072500171\\
					0.1	0.530625072500171\\
					0.1	0.525840840454514\\
					0.1	0.521648337402094\\
					0.1	0.521252985532249\\
					0.1	0.521252985532249\\
					0.1	0.514013722385443\\
					0.1	0.514013722385443\\
					0.1	0.508381451982377\\
					0.1	0.508381451982377\\
					0.1	0.508381451982377\\
					0.1	0.508381451982377\\
					0.1	0.50797644418321\\
					0.1	0.50797644418321\\
					0.1	0.493346042732215\\
					0.1	0.493346042732215\\
					0.1	0.485901428140046\\
					0.1	0.485901428140046\\
					0.1	0.485512557083268\\
					0.1	0.485512557083268\\
					0.1	0.485512557083268\\
					0.1	0.48517199001783\\
					0.1	0.48517199001783\\
					0.1	0.481980251949233\\
					0.1	0.481980251949233\\
					0.1	0.479583370111743\\
					0.1	0.479583370111743\\
					0.1	0.472923537189286\\
					0.1	0.472367399244587\\
					0.1	0.470099827453861\\
					0.1	0.470099827453861\\
					0.1	0.468449220044945\\
					0.1	0.464920367838359\\
					0.1	0.464920367838359\\
					0.1	0.464920367838358\\
					0.1	0.464920367838358\\
					0.1	0.457713600794981\\
					0.1	0.457713600794981\\
					0.1	0.457713600794981\\
					0.1	0.45771360079498\\
					0.1	0.456341752225864\\
					0.1	0.454685727107353\\
					0.1	0.454685727107353\\
					0.1	0.454685727107352\\
					0.1	0.452208026553418\\
					0.1	0.449420928264365\\
					0.1	0.434570321810979\\
					0.1	0.433361574151994\\
					0.1	0.425766952498131\\
					0.1	0.425766952498131\\
					0.1	0.424493620237877\\
					0.1	0.420309560395941\\
					0.1	0.419601344085339\\
					0.1	0.419601344085339\\
					0.1	0.419601344085339\\
					0.1	0.419601344085339\\
					0.1	0.416386752154004\\
					0.1	0.416386752154004\\
					0.1	0.416386752154004\\
					0.1	0.416386752154004\\
					0.1	0.413676068660746\\
					0.1	0.413676068660745\\
					0.1	0.413676068660745\\
					0.1	0.396773244377371\\
					0.1	0.396773244377371\\
					0.1	0.393799611114702\\
					0.1	0.390226314011638\\
					0.1	0.390226314011638\\
					0.1	0.384277368566017\\
					0.1	0.382675218253471\\
					0.1	0.382675218253471\\
					0.1	0.382675218253471\\
					0.1	0.376958278263917\\
					0.1	0.375650500670558\\
					0.1	0.375650500670558\\
					0.1	0.375217289721437\\
					0.1	0.375059903513028\\
					0.1	0.375059903513028\\
					0.1	0.375059903513028\\
					0.1	0.375059903513028\\
					0.1	0.369529023390588\\
					0.1	0.361735285398626\\
					0.1	0.361735285398626\\
					0.1	0.360169898238937\\
					0.1	0.358976166943764\\
					0.1	0.358976166943764\\
					0.1	0.355959007785398\\
					0.1	0.35423764807741\\
					0.1	0.348245176736094\\
					0.1	0.343148080048003\\
					0.1	0.343148080048003\\
					0.1	0.343148080048003\\
					0.1	0.33958348400881\\
					0.1	0.33958348400881\\
					0.1	0.33958348400881\\
					0.1	0.338763928120032\\
					0.1	0.336994334363306\\
					0.1	0.336994334363306\\
					0.1	0.333733054872052\\
					0.1	0.333733054872052\\
					0.1	0.333733054872052\\
					0.1	0.333733054872052\\
					0.1	0.333245604133161\\
					0.1	0.330360669873092\\
					0.1	0.330360669873092\\
					0.1	0.321691701559159\\
					0.1	0.321179089510156\\
					0.1	0.321179089510156\\
					0.1	0.320064617208594\\
					0.1	0.318423331479471\\
					0.1	0.318423331479471\\
					0.1	0.31221298490617\\
					0.1	0.31221298490617\\
					0.1	0.310645659425448\\
					0.1	0.310645659425448\\
					0.1	0.310645659425448\\
					0.1	0.307276605986977\\
					0.1	0.307276605986977\\
					0.1	0.307276605986977\\
					0.1	0.302711493656091\\
					0.1	0.302621845992337\\
					0.1	0.302621845992337\\
					0.1	0.300569577976147\\
					0.1	0.296962184875733\\
					0.1	0.279959336178251\\
					0.1	0.279959336178251\\
					0.1	0.276539070968107\\
					0.1	0.276539070968107\\
					0.1	0.276539070968106\\
					0.1	0.276407119562671\\
					0.1	0.27640711956267\\
					0.1	0.276152414357214\\
					0.1	0.268249357621367\\
					0.1	0.268249357621367\\
					0.1	0.262375227832262\\
					0.1	0.260678765618306\\
					0.1	0.249201483058166\\
					0.1	0.249201483058166\\
					0.1	0.249201483058166\\
					0.1	0.249201483058166\\
					0.1	0.243690856872702\\
					0.1	0.243690856872702\\
					0.1	0.243690856872702\\
					0.1	0.239854055147908\\
					0.1	0.239854055147908\\
					0.1	0.23439090764587\\
					0.1	0.23439090764587\\
					0.1	0.233876869250398\\
					0.1	0.233876869250398\\
					0.1	0.224395346360879\\
					0.1	0.224395346360879\\
					0.1	0.224180877688377\\
					0.1	0.213720990418992\\
					0.1	0.213720990418992\\
					0.1	0.211229299388191\\
					0.1	0.211229299388191\\
					0.1	0.211229299388191\\
					0.1	0.205274340254908\\
					0.1	0.205274340254908\\
					0.1	0.205274340254907\\
					0.1	0.205274340254907\\
					0.1	0.199748774117565\\
					0.1	0.199504380879429\\
					0.1	0.199504380879429\\
					0.1	0.192374695729069\\
					0.1	0.192374695729069\\
					0.1	0.188111927103452\\
					0.1	0.188111927103452\\
					0.1	0.185986527544492\\
					0.1	0.178767741903679\\
					0.1	0.178767741903679\\
					0.1	0.178767741903679\\
					0.1	0.178494287245091\\
					0.1	0.178494287245091\\
					0.1	0.178494287245091\\
					0.1	0.165131892508459\\
					0.1	0.165131892508459\\
					0.1	0.161347197451649\\
					0.1	0.161347197451649\\
					0.1	0.161347197451649\\
					0.1	0.161347197451649\\
					0.1	0.159643493087222\\
					0.1	0.159643493087222\\
					0.1	0.151828507846025\\
					0.1	0.150358483812268\\
					0.1	0.150358483812268\\
					0.1	0.147943660647037\\
					0.1	0.147943660647037\\
					0.1	0.147943660647037\\
					0.1	0.147792177400607\\
					0.1	0.147119954868653\\
					0.1	0.147119954868653\\
					0.1	0.147119954868653\\
					0.1	0.146306184419168\\
					0.1	0.146306184419168\\
					0.1	0.13075940413749\\
					0.1	0.13075940413749\\
					0.1	0.119538212056879\\
					0.1	0.119538212056879\\
					0.1	0.11742005464839\\
					0.1	0.11742005464839\\
					0.1	0.11742005464839\\
					0.1	0.11742005464839\\
					0.1	0.117393034048984\\
					0.1	0.117393034048984\\
					0.1	0.117393034048984\\
					0.1	0.115545088588597\\
					0.1	0.113844626934656\\
					0.1	0.113844626934656\\
					0.1	0.109597827256721\\
					0.1	0.108342271895467\\
					0.1	0.108342271895467\\
					0.1	0.101281881178936\\
					0.1	0.101281881178936\\
					0.1	0.101281881178936\\
					0.1	0.0963869157665208\\
					0.1	0.0963869157665207\\
					0.1	0.08684240745093\\
					0.1	0.0868424074509299\\
					0.1	0.081383069450145\\
					0.1	0.081383069450145\\
					0.1	0.081383069450145\\
					0.1	0.0794329310265362\\
					0.1	0.0794329310265361\\
					0.1	0.0792616693311703\\
					0.1	0.0792616693311702\\
					0.1	0.0734929118451314\\
					0.1	0.0734929118451313\\
					0.1	0.0734929118451313\\
					0.1	0.0734929118451312\\
					0.1	0.0714034771128363\\
					0.1	0.0663260599786661\\
					0.1	0.0663260599786661\\
					0.1	0.0620144273955514\\
					0.1	0.0620144273955514\\
					0.1	0.0562917808528765\\
					0.1	0.0562917808528764\\
					0.1	0.0554438074892196\\
					0.1	0.0554438074892196\\
					0.1	0.0554438074892195\\
					0.1	0.0489215119656336\\
					0.1	0.0489215119656335\\
					0.1	0.042978250073743\\
					0.1	0.042978250073743\\
					0.1	0.0393276499961932\\
					0.1	0.0393276499961931\\
					0.1	0.0332091269689512\\
					0.1	0.0295657690418726\\
					0.1	0.0295657690418725\\
					0.1	0.0295657690418724\\
					0.1	0.0276419390245821\\
					0.1	0.027641939024582\\
					0.1	0.0257411542548229\\
					0.1	0.0257411542548228\\
					0.1	0.0243098480618653\\
					0.1	0.0243098480618652\\
					0.1	0.0164599544811221\\
					0.1	0.016459954481122\\
					0.1	0.009605733799503\\
					0.1	0.00960573379950298\\
					0.1	0.00960573379950292\\
					0.1	0.00669483081631583\\
					0.1	0.00669483081631578\\
					0.1	0\\
				};
				\addlegendentry{Theorem~\ref{cor:MAC:Mari}}
				
			\end{axis}
			
			%
			%
			
		\end{tikzpicture}%
		\caption{Sum-rate distortion tradeoff achieved by Theorems~\ref{cor:MAC:Mari}   and~\ref{Th:achievability:MAC} in Example~\ref{Ex:MAC:NoisyFB} for given channel parameters $p_s=0.9$, $t_0=0.3$, $t_1=0.1$ and $t_2=0.1$.}	\label{fig:plot}
		
	\end{figure}
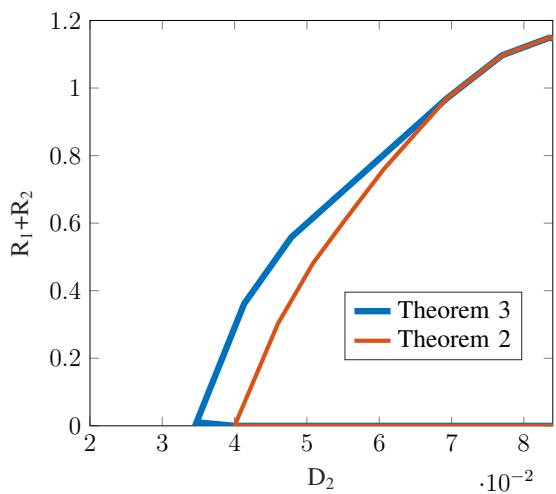
	%
	For channel parameters $p_s=0.9$, $t_0=0.3$, $t_1=0.1$ and $t_2=0.1$ and above choices of random variables, Figure~\ref{fig:plot} shows the maximum sum-rate $R_1+R_2$ in function of distortion $\D_2$ achieved by Theorems~\ref{Th:achievability:MAC} 
	and \ref{cor:MAC:Mari}, where recall that for the region in Theorem~\ref{cor:MAC:Mari} we set  $V_1=V_2=0$. Notice that both curves are strictly concave and thus improve over classic time- and resource sharing strategies. The minimum distortions achieved by Theorems~\ref{Th:achievability:MAC} and \ref{cor:MAC:Mari}  are  $D_{2,\min}=0.035$  and  $D_{2,\min}=0.04$.

	\section{Device-to-Device Communication (The Two-Way Channel)}\label{sec:D2D_model}
	
	In this section, we consider the  ISAC two-way channel, where two terminals exchange data over a common channel and based on their inputs and outputs also wish to estimate the state-sequences that govern the two-way channel. 
	
	
	\subsection{System Model}

\begin{figure}[h!]
		\centering
		\includegraphics[width=0.85\textwidth]{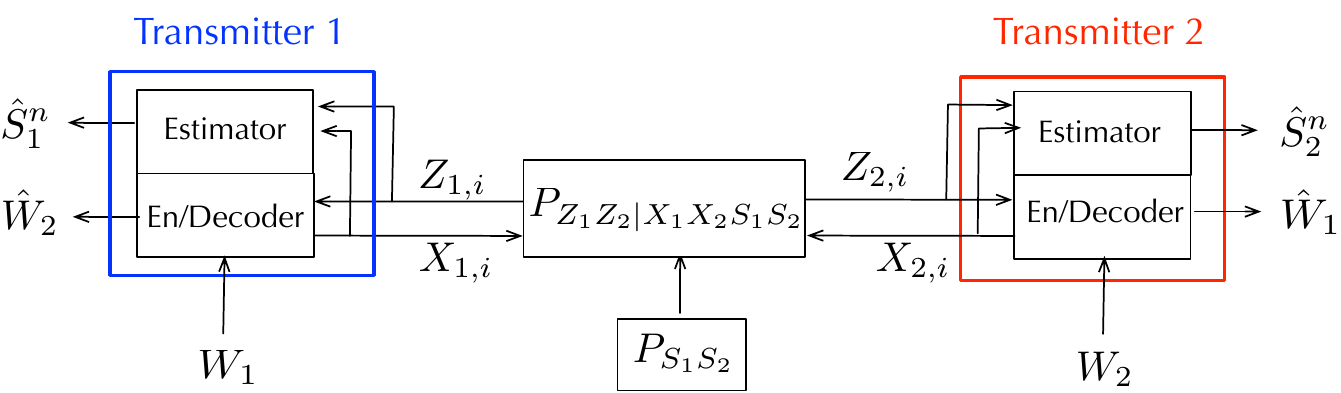}
		
		
		\caption{State-dependent discrete memoryless two-way channel with sensing at the terminals.}
		\label{fig:ModelD2D}
		\hspace{-10cm}
	\end{figure}

	Consider the two-terminal two-way communication scenario  in Fig.~\ref{fig:ModelD2D}.  The model consists of a two-dimensional memoryless state sequence $\{(S_{1, i},  S_{2, i})\}_{i\geq 1}$ whose samples at any given time $i$ are distributed according to a given joint law $P_{S_1S_2}$ over the state alphabets $\Sc_1\times \Sc_2$. 
	Given that at time-$i$ Tx 1 sends input $X_{1,i}=x_1$ and Tx~2 input $X_{2,i}=x_2$ and given  state realizations $S_{1,i}=s_{1}$ and $S_{2,i}=s_{2}$,  the Txs' time-$i$  feedback signals $Z_{1,i}$ and $Z_{2,i}$ are distributed according to the stationary channel transition law $ P_{Z_1Z_2|S_1S_2X_1X_2}(\cdot,\cdot|s_1,s_2,x_1,x_2)$. Input and output 
	alphabets $\Xc_1, \Xc_2,  \Yc,  \Zc_1,  \Zc_2, \Sc_1, \Sc_2$ are assumed  finite.\footnote{\textcolor{black}{The results can be extended to well-behaved continuous channels.}}

	A $(2^{n\R_1}, 2^{n\R_2},  n)$ code  consists  of
	\begin{enumerate}
		\item two message sets $\Wc_1= [1:2^{n\R_1}]$ and $\Wc_2= [1:2^{n\R_2}]$;
		\item sequences of encoding functions $\Omega_{k,i}\colon \Wc_k \times \Zc_k^{{i-1} }\to \Xc_k$,  for $i=1, 2, \ldots, n$ and $k=1,2$; 
		\item  decoding functions $g_k\colon \Zc^n \to \Wc_k$, for $k=1,2$;
		\item  state estimators  $\phi_k \colon \Xc_k^n \times \Zc_k^n \to \hat{\Sc}_k^n$,   for $k=1,2$, where  $\hat{\Sc}_1$ and $\hat{\Sc}_2$ are given  reconstruction alphabets.
	\end{enumerate}

	
	\textcolor{black}{Fix a blocklength $n$, rates $R_1, R_2\geq 0$, and a $(2^{nR_1},2^{nR_2},n)$-code $(\{\Omega_{1,i}\},\{\Omega_{2,i}\},  g_1,g_2, \phi_1,\phi_2)$. Let then the random message $W_k$ be uniformly distributed over the message set $\Wc_k$, for each $k=1,2$, and  generate the inputs  according to the encoding function $X_{k,i}=\Omega_{k,i}(W_k, Z_{k}^{i-1})$,  for $i=1, \ldots,  n$.} 
	Tx~$k\in\{1,2\}$ obtains its state estimate 
	as $\hat{S}_k^n:= (\hat{S}_{k, 1}, \cdots, \hat{S}_{k, n} )=\phi_k(X_k^n,  Z_k^n)$ and its message  guess as  $ \hat{W}_{3-k}=g_k(Z_k^n,W_k)$
	
	We shall measure the quality of the state estimates $\hat{S}_k^n$   by  bounded per-symbol distortion functions $d_k\colon \Sc_k\times \hat{\Sc}_k \mapsto [0, \infty)$, 
	and consider \emph{expected average block distortions}
	\begin{equation}
		\Delta_k^{(n)}:= \frac{1}{n} \sum_{i=1}^n \mathbb{E}\left[d_k\left(S_{k, i},  \hat{S}_{k, i}\right)\right],  \quad k=1, 2.
	\end{equation}
	The probability of decoding error is defined as:
	\begin{IEEEeqnarray}{rCl}
		P^{(n)}_e& := &\textnormal{Pr}\Big( \hat{W}_1 \neq W_1 
		\quad \textnormal{or} 
		\quad \hat{W}_2\neq W_2 \Big).
	\end{IEEEeqnarray}
	\begin{definition} 
		A rate-distortion tuple $(\R_1,  \R_2,  \D_1,  \D_2)$ is
		achievable if there exists  a sequence (in $n$) of  $(2^{n\R_1}, 2^{n\R_2},  n)$ codes that simultaneously satisfy
		\begin{subequations}\label{eq:asymptotics}
			\begin{IEEEeqnarray}{rCl}
				\lim_{n\to \infty}	P^{(n)}_e 
				&=&0 \\
				\varlimsup_{n\to \infty}	\Delta_k^{(n)}& \leq& \D_k,  \quad \textnormal{for } k=1, 2.\label{eq:asymptotics_dis}
			\end{IEEEeqnarray}
		\end{subequations}
	\end{definition}

	\begin{definition}
		The capacity-distortion region $\CDc$ is the closure of the set of all achievable tuples $(\R_1,  \R_2, \D_1, \D_2)$.
	\end{definition}

	\begin{remark}[State-Information at the Terminals]\label{rem:states2}
		Considering a two-way channel where 
		\begin{equation}\label{eq:state_info}
			Z_k=(S_{\bar{k}},Z_k'), \qquad k\in\{1,2\},
		\end{equation}
		for some output $Z_k'$. This models a situation where each terminal  obtains strictly causal state-information about the other terminal's state. Inner bounds for this setup with strictly causal state-information can immediately be obtained from our results presented in the next section by plugging in the choice in \eqref{eq:state_info}. 
		The same remark applies also to imperfect strictly-causal state-information in which case the output should be modelled as 
		\begin{equation}\label{eq:state_info2}
			Z_k=(T_k,Z_k'), \qquad k\in\{1,2\},
		\end{equation}
		where $Z_{k}'$ again models the actual channel output and  $T_k$ models the strictly causal imperfect state-information at Terminal $k$. Alternatively, $T_k$ could even be related to the desired channel state $S_k$ and not only to the other terminal's state $S_{\bar{k}}$. Plugging  the choice \eqref{eq:state_info2} into our results for an appropriate choice of $T_k$ leads to results for this related setup with imperfect or generalized state-information at the terminals. 
		
		In contrast,  our model does not include  causal or non-causal state-information. These  are interesting extensions of our work, but left for future research. They would certainly require new tools such as dirty-paper coding \cite{DPC}. 
	\end{remark} 
	
	
	\section{A Collaborative ISAC Scheme for Device-to-Device Communication}\label{sec:D2D_scheme}
	
	
	We first review Han's scheme for pure data communication over the two-way channel and then include the collaborative sensing idea in Han's scheme. Finally we integrate  collaborative sensing  and communication through  joint source-channel coding (JSCC).

	\subsection{Han's Two-Way Coding Scheme}
	\begin{figure}[h!]
		\centering
		\includegraphics[width=\textwidth]{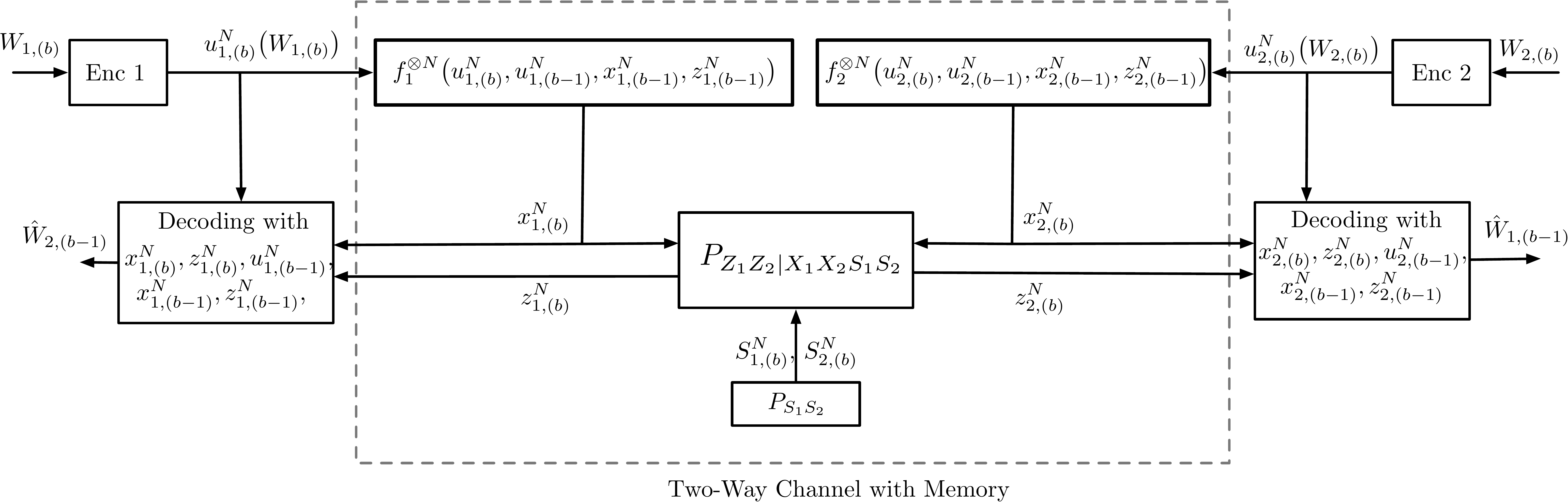}
		\caption{Han's coding scheme in a given block $b$. Encoders transform the discrete-memoryless two-way channel into a channel with memory so as to be able to correlate the inputs of the two terminals. Encoding is then performed through the independent codewords $u_{1,(b)}^N$ and $u_{2,(b)}^N$. Decoding of block-$(b-1)$ messages is performed based on the inputs/outputs in the two consecutive blocks $b-1$ and $b$. }
		\label{fig:TW1}
		\hspace{-10cm}
	\end{figure}
	
	\textcolor{black}{The capacity region of the two-way channel, and thus the optimal coding scheme is still open for general channels. Various inner and outer bounds  on the capacity region have been proposed. For example, Schalkwijk proposed an interesting inner bound  for a binary multiplier channel based on a method that iteratively describes message points, an approach that is reminiscent of single-user feedback schemes. Han \cite{Han84} and   Kramer \cite{Kramer:thesis} proposed schemes that  correlate  the inputs of the two terminals in a block-fashion. While for Han's coding scheme the correlation ensures a stationary distribution of the inputs and outputs across the blocks and thus still allows for single-letter rate-expressions, Kramer has to resort to multi-letter rate-expressions based on directed mutual informations. 
		An interesting outer bound on the capacity region was proposed by Hekstra and Willems \cite{Hekstra89} again based on the dependence-balance idea, similar to the MAC with feedback. }
	
	\textcolor{black}{The ISAC scheme we present in this manuscript is based on Han's  coding scheme, which is depicted in Figure~\ref{fig:TW1} and described in the following.}
	For convenience of notation, define
	\begin{equation}
		P_{Z_1Z_2|X_1X_2}(z_1,z_2|x_1,x_2) = \sum_{s_1 \in \mathcal{S}_1, s_{2}\in\mathcal{S}_2} P_{S_1S_2}(s_1,s_2) P_{Z_1Z_2|X_1X_2S_1S_2}(z_1,z_2|x_1,x_2,s_1,s_2) .
	\end{equation} 
	
	Han's scheme splits the blocklength $n$ into $B+1$ blocks of length $N=n/(B+1)$ each.  Accordingly, throughout, we let  $X_{1,(b)}^N, X_{2,(b)}^N, S_{1,(b)}^N, S_{2,(b)}^N,  Z_{1,(b)}^N, Z_{2,(b)}^N$ denote the block-$b$ inputs, states and outputs, e.g., $S_{1,(b)}^N:=(S_{1(b-1)N+1}, \ldots, S_{1, bN})$. We also represent the two messages $W_1$ and $W_2$ in a one-to-one way as the $B$-length tuples 
	\begin{equation}
		W_{k}=(W_{k,(1)},\ldots, W_{k,(B)}), \qquad k\in\{1,2\}, 
	\end{equation}
	where each $W_{k,(b)}$ is independent and uniformly distributed over $\left[2^{N\bar {R}_k}\right]$ for $\bar{R}_k\triangleq \frac{B+1}{B}R_k$.
	
	Construct an independent code $\mathcal{C}_{k,(b)}=\Big\{ u_{k,(b)}^N(1), \ldots, u_{k,(b)}^N\big(2^{n \bar{R}_k}\big)\Big\}$ for each of the two terminals by picking entries i.i.d. according to some pmf $P_{U_{k}}$. As shown in Figure~\ref{fig:TW1}, Terminal $k$ encodes Message $W_{k,(b)}$ by means of the codeword $u_{k,(b)}^{N}(W_{k,(b)})$ and sends the sequence 
	\begin{equation}
		X_{k,(b)}^N= f_{k}^{\otimes N} \left( u_{k,(b)}^{N}(W_{k,(b)}), \ u_{k,(b-1)}^{N}(W_{k,(b-1)}), \  x_{k, (b-1)}^N, \ z_{k,(b-1)}^N  \right)
	\end{equation}
	over the channel during block $b$. Notice that by applying the function $f_k$ to the block-$b$ codeword symbols as well as to the symbols of the block-$(b-1)$  codeword $u_{k,(b-1)}^N(W_{k,(b-1)})$ and the block-$(b-1)$ channel inputs  and outputs $x_{k,(b-1)}^N$ and $z_{k,(b-1)}^N$,  the terminals introduce memory to the channel. An interesting point of view is to consider the transition of the codewords $u_{1,(b)}^N$ and $u_{2,(b)}^N$ to the channel outputs $z_{1,(b)}^N$ and $z_{2,(b)}^N$ as a virtual two-way channel with block-memory over which one can code and decode. Naturally, decoding of each message part $W_{k,(b)}$  is not based only on the signals in  block $(b)$ because other blocks depend on this message as well. In Han's scheme, decoding is over two consecutive blocks. Specifically, Terminal $k$ decodes the   block-$b$ message $W_{\bar{k},(b)}$ using a joint-typicality decoder  based on the block-$b$ inputs, outputs, and own transmitted codewords $ x_{k,(b)}^N, z_{k,(b)}^N$ and $u_{k,(b)}^N$, as well as on the block-$(b+1)$ inputs and outputs $x_{k,(b+1)}^N$ and $z_{k,(b+1)}^N$. 
	
	Notice that without any special care, the rate-region that is achievable with above scheme has to be described with a multi-letter expression because the joint pmf of the tuple $x_{1,(b+1)}^N, z_{1,(b+1)}^N, u_{1,(b)}^N, x_{1,(b)}^N, z_{1,(b)}^N$ that Terminal~1 uses to decode codeword $u_{2,(b)}^N(W_{k,(b)})$ varies with the block $b$. However,  if one  chooses a joint pmf $P_{U_1U_2X_1X_2Z_1Z_2}$ satisfying the stationarity condition
	\begin{IEEEeqnarray}{rCl}
		\lefteqn{P_{{U}_1{U}_2X_1X_2Z_1Z_2}(u_1,u_2, x_1,x_2,z_1,z_2)}\quad \nonumber \\
		&=& \sum_{\tilde{u}_1,\tilde{u}_2, \tilde{x}_1,\tilde{x}_2,\tilde{z}_1,\tilde{z}_2} P_{Z_1Z_2|X_1X_2}(z_1,z_2|x_1,x_2)		\mathbbm{1}\{x_1=f_1(u_1,\tilde{u}_1,\tilde{x}_1,\tilde{z}_1)\} 
		\nonumber \\
		&&
		\mathbbm{1}\{x_2=f_2(u_2,\tilde{u}_2,\tilde{x}_2, \tilde{z}_2)\} \cdot P_{U_1}(u_1)P_{U_2}(u_2) P_{{U}_1{U}_2{X}_1{X}_2Z_1Z_2}(\tilde{u}_1,\tilde{u}_2,\tilde{x}_1,\tilde{x}_2,\tilde{z}_1,\tilde{z}_2), \label{eq:stationarity}
	\end{IEEEeqnarray}	
	where  $P_{U_1}$ and $P_{U_2}$ are the marginals of $P_{U_1U_2X_1X_2Z_1Z_2}$, then 
	the pmf of the tuple  of sequences $x_{1,(b+1)}^N, x\textcolor{black}{^N}_{2,(b+1)}, z_{1,(b+1)}^N, z_{2,(b+1)}^N, u_{1,(b)}^N, u_{2,(b)}^N, x_{1,(b)}^N, x_{2,(b)}^N, z_{1,(b)}^N, z_{2,(b)}^N$ is independent of the block index $b$. This allows to characterize the rate region achieved by the described coding scheme using a  single-letter expression.  All rate-pairs $(R_1,R_2)$  are achievable that satisfy 
	\begin{subequations}\label{eq:Han}
		\begin{IEEEeqnarray}{rCl} 
			R_1&\leq&  I(U_1;  X_2,Z_2, \tilde{U}_2, \tilde{X}_2,\tilde{Z}_2)\\
			R_2&\leq & I(U_2; X_1,Z_1, \tilde{U}_1, \tilde{X}_1,\tilde{Z}_1),
		\end{IEEEeqnarray}
	\end{subequations}
	where $(U_1,U_2,X_1,X_2,Z_1,Z_2,\tilde{U}_1, \tilde{U}_2,\tilde{X}_1,\tilde{X}_2,\tilde{Z}_1 ,\tilde{Z}_2)$ are distributed according to the pmf
	\begin{IEEEeqnarray}{rCl}
		\lefteqn{P_{U_1U_2X_1X_2Z_1Z_2\tilde{U}_1 \tilde{U}_2\tilde{X}_1\tilde{X}_2\tilde{Z}_1 \tilde{Z}_2}(u_1,u_2,x_1,x_2,z_1,z_2,\tilde{u}_1, \tilde{u}_2,\tilde{x}_1,\tilde{x}_2,\tilde{z}_1 ,\tilde{z}_2)} \quad \nonumber \\
		& =& P_{Z_1Z_2|X_1X_2}(z_1,z_2|x_1,x_2)		\mathbbm{1}\{x_1=f_1(u_1,\tilde{u}_1,\tilde{x}_1,\tilde{z}_1)\} \mathbbm{1}\{x_2=f_2(u_2,\tilde{u}_2,\tilde{x}_2, \tilde{z}_2)\}\nonumber \\
		&& \cdot P_{U_1}(u_1)P_{U_2}(u_2) P_{{U}_1{U}_2{X}_1{X}_2Z_1Z_2}(\tilde{u}_1,\tilde{u}_2,\tilde{x}_1,\tilde{x}_2,\tilde{z}_1,\tilde{z}_2).\label{eq:distribution}
	\end{IEEEeqnarray}	
	This recovers Han's theorem:
	\begin{theorem}[Han's Achievable Region for Two-Way Channels \cite{Han84}]\label{thm:Han}
		Any nonnegative rate-pair $(R_1,R_2)$ is achievable over the two-way channel if it  satisfies Inequalities \eqref{eq:Han} for some choice of  pmf $P_{U_1U_2X_1X_2Z_1Z_2}$  and functions $f_1$ and $f_2$ satisfying the stationarity condition \eqref{eq:stationarity}.
	\end{theorem}
	
	For certain cases \textcolor{black}{the} above theorem can be simplified, and for certain channels the simplified  region even coincides with  capacity. The simplification is obtained by choosing  the two functions $f_1$ and $f_2$ to simply produce the  codewords $u_{1,(b-1)}^N$ and $u_{2,(b-1)}^N$ from the previous block\footnote{The delay of a block introduced in this scheme is not crucial, it simply comes from the fact that Han's scheme decodes the block-$(b-1)$ codewords based on the block-$b$ outputs. In this special case without adaptation, Han's scheme could be simplified by transmitting and decoding the codewords $u_{1,(b-1)}^N$ and $u_{2,(b-1)}^N$ directly in block $b-1$ without further delay.} and ignore the other arguments. In this case, the set of rates that can be achieved coincides with the following inner bound that was first proposed by Shannon \cite{Shannon93Twoway}.
	
	\begin{theorem}[Shannon's Inner Bound, \cite{Shannon93Twoway}]\label{thm:Shannon}
		A pair of  nonnegative pairs $(R_1,R_2)$ is achievable if it satisfies
		\begin{subequations}\label{eq:Shannon}
			\begin{IEEEeqnarray}{rCl} 
				R_1\leq  I(X_1; Z_2|X_2)\\
				R_2\leq  I(X_2; Z_1|\textcolor{black}{X_1}),
			\end{IEEEeqnarray}
		\end{subequations}
		for some input pmfs $P_{X_1}$ and $P_{X_2}$ and where $(X_1,X_2,Z_1,Z_2)\sim P_{X_1}P_{X_2}P_{Z_1Z_2|X_1X_2}$. 
	\end{theorem}

	\subsection{Collaborative Sensing and Communication  based on Han's Two-Way Coding Scheme}
	We extend Han's coding scheme to include also collaborative sensing, that means each terminal compresses its block-$b$ inputs and outputs so as to capture information about the other terminal's state and sends this state-information in the next-following block. In this first collaborative sensing and communication scheme that we present here, the sensing (compression) does not affect the communication (except possibly for the choice of the pmf $P_{U_1U_2X_1X_2Z_1Z_2}$). 
	\begin{figure}[h!]
		\centering
		\includegraphics[scale=0.35]{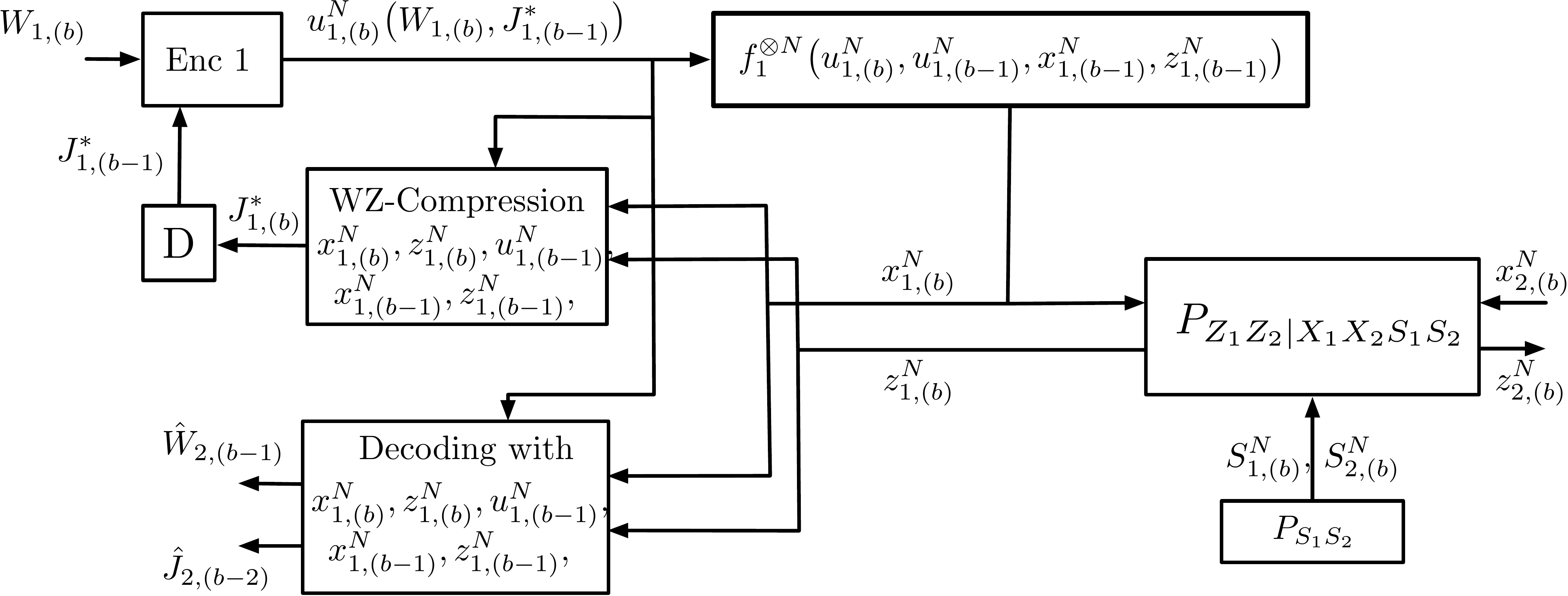}
		\caption{A first collaborative-sensing  version of Han's coding scheme. The figure illustrates the encoding and decoding operations in a given block $b$ at Terminal 1; Terminal 2 behaves analogously.  To facilitate sensing at Terminal 2, Terminal 1 compresses its block-$b$ channel inputs and outputs, together with its inputs, outputs, and codeword from the previous block $(b-1)$ (which are all resent in block $b$) using Wyner-Ziv compression \cite{WynerZiv76} to account for the side-information at Terminal 2.}
		\label{fig:TW2}
		\hspace{-10cm}
	\end{figure}
	In fact, we again use Han's encodings and decodings as described in the previous subsection, except that the block-$b$ codeword not only encodes message $W_{k,(b)}$ but also a  compression  index  $J_{k,(b-1)}^*$ that  carries information about the block-$(b-1)$ state $S_{\bar{k},(b-1)}$. This compression index is then decoded at Terminal $\bar{k}$ after block $(b+1)$ simultaneously with message $W_{k,(b)}$.   See Figure~\ref{fig:TW2}.
	
	The analysis of the communication-part of our ISAC scheme is similar  as in Han's scheme. Since  the  compression indices take parts of the place reserved for  ordinary messages in Han's scheme,  their rates $R_{\textnormal{WZ},1}$ and $R_{\textnormal{WZ},2}$ have to be subtracted from Han's communication rates. We thus have the following constraints for reliable communication and reliable decoding of the compression indices: 
	\begin{subequations}\label{eq:Rates_minus}
		\begin{IEEEeqnarray}{rCl} 
			R_1+  R_{\textnormal{WZ},1}& \leq & I(U_1;  X_2,Z_2, \tilde{U}_2, \tilde{X}_2,\tilde{Z}_2)  \\
			R_2 +R_{\textnormal{WZ},2}& \leq & I(U_2; X_1,Z_1, \tilde{U}_1, \tilde{X}_1,\tilde{Z}_1).
		\end{IEEEeqnarray}
	\end{subequations}

	It remains to  explain the compression and state estimation in more details. In our scheme, the index $J_{k,(b-1)}^*$ is obtained by means of a  Wyner-Ziv compression  \cite{WynerZiv76} that lossily compresses the tuple $(x_{k,(b-1)}^N, z_{k,(b-1)}^N, u_{k,(b-2)}^N, x_{k,(b-2)}^N, z_{k,(b-2)}^N)$ for a decoder that has side-information $(x_{\bar{k},(b-1)}^N, z_{\bar{k},(b-1)}^N, u_{\bar{k},(b-2)}^N, x_{\bar{k},(b-2)}^N, z_{\bar{k},(b-2)}^N)$. In order for the decoder to be able to correctly reconstruct the compression codeword, the  Wyner-Ziv codes  need to be of rates at least \cite{WynerZiv76}
	\begin{IEEEeqnarray}{rCl}\label{eq:WZ}
		R_{\textnormal{WZ},k} > I(V_k ;  X_k, Z_k, \tilde{U}_k, \tilde{X}_k, \tilde{Z}_k | X_{\bar{k}}, Z_{\bar{k}}, \tilde{U}_{\bar{k}}, \tilde{X}_{\bar{k}} ,\tilde{Z}_{\bar{k}}), \qquad k\in\{1,2\},
	\end{IEEEeqnarray} 
	where the tuple  $(U_1,U_2,X_1,X_2,Z_1,Z_2,\tilde{U}_1, \tilde{U}_2,\tilde{X}_1,\tilde{X}_2,\tilde{Z}_1 ,\tilde{Z}_2)$ refers to the auxiliary random variables  chosen by Han's scheme of joint pmf  as in \eqref{eq:distribution} and $V_1$ and $V_2$ can be any random variables satisfying the Markov chains: 
	\begin{IEEEeqnarray}{rCl}
		V_k  \to ( X_k, Z_k, \tilde{U}_k, \tilde{X}_k ,\tilde{Z}_k ) \to (X_{\bar{k}}, Z_{\bar{k}}, \tilde{U}_{\bar{k}}, \tilde{X}_{\bar{k}} ,\tilde{Z}_{\bar{k}}, S_{k}, S_{\bar{k}}).  \label{eq:Markov1}
	\end{IEEEeqnarray}
	In Wyner-Ziv coding, the encoder produces a codeword that is then reconstructed also at the receiver. We shall denote these codewords by $v_{k,(b-1)}^N(J_{k,(b-1)}^*, \ell_{k,(b-1)})$, for $k\in\{1,2\}$, where  $\ell_{k,(b-1)}$ denotes a binning-index that does not have to be conveyed to the Terminal $\bar{k}$  because this latter can recover it from its side-information. Thus, after block $(b+1)$ and after decoding index $J_{{k},(b-1)}^*$,   with high probability Terminal $\bar{k}$  can reconstruct the  codeword $v_{{k},(b-1)}^N(J_{k,(b-1)}^*, \ell_{k,(b-1)})$ chosen at Terminal $k$.  
	
	Terminal $k$ can wait arbitrarily long to produce an estimate of the state-sequence $S_k^N$. We propose that it waits after the block-$(b+1)$ decoding to  reconstruct the block-$b$ state $S_{k,(b)}^N$ by applying an optimal symbol-by-symbol estimator  to the related sequences of inputs, outputs, and channel codewords of blocks $b-1$ and $b$, as well as on the compression codeword $v_{\bar{k},(b)}^N$: 
	\begin{equation}\label{eq:state_estimate1}
		\hat{S}^N_{k,(b)} = \tilde{\phi}_{2,k}^{*\otimes N} \left(  v_{\bar{k},(b)}^N, x_{k,(b)}^N,  z_{k,(b)}^N, \hat{u}_{\bar{k},(b)}, u_{k,(b-1)}^N, x_{k,(b-1)}^N,  z_{k,(b-1)}^N,\hat{u}_{\bar{k},(b-1)} \right),
	\end{equation}
	where 
	\begin{IEEEeqnarray}{rCl}
		\tilde{\phi}_{2,k}^*(v_{\bar{k}}, x_k,z_k,u_{\bar{k}}, \tilde{u}_k, \tilde{x}_{k}, \tilde{z}_k, \tilde{u}_{\bar{k}}):= 
		\textnormal{arg}\min_{s_k'\in \hat{\mathcal{S}_k}} \sum_{s_k\in \mathcal{S}_k}  P_{S_k|X_kZ_kU_{\bar{k}}}(s_k|x_k,z_k,u_{\bar{k}})\;  d_k(s_k,  s_k').
	\end{IEEEeqnarray}

	By \eqref{eq:Rates_minus} and \eqref{eq:WZ} and standard typicality arguments, one obtains the following theorem. (The theorem is a special case of the next-following theorem, for which we provide a detailed analysis in the extended version \cite{ArxivMacIt}.)
	\begin{theorem}[Inner Bound via Separate Source-Channel Coding]\label{thm:Sep}
		Any nonnegative rate-distortion quadruple $(R_1,R_2, D_1,D_2)$ is achievable if it  satisfies  the following two rate-constraints  
		\begin{subequations}\label{eq:HanCompression}
			\begin{IEEEeqnarray}{rCl} 
				R_1 & \leq & I(U_1;  X_2,Z_2, \tilde{U}_2, \tilde{X}_2,\tilde{Z}_2) - I(V_1 ;  X_1, Z_1, \tilde{U}_1, \tilde{X}_1 ,\tilde{Z}_1 | X_{2}, Z_2, \tilde{U}_2, \tilde{X}_2 ,\tilde{Z}_2) \\
				R_2 & \leq & I(U_2; X_1,Z_1, \tilde{U}_1, \tilde{X}_1,\tilde{Z}_1)- I(V_2 ;  X_2, Z_2, \tilde{U}_2, \tilde{X}_2 ,\tilde{Z}_2 | X_{1}, Z_1, \tilde{U}_1, \tilde{X}_1, \tilde{Z}_1),
			\end{IEEEeqnarray}
			and the two distortion constraints
			\begin{IEEEeqnarray}{rCl} 
				\E{ d_1\big( S_1,\ \tilde{\phi}_{2,1}^*( V_2, X_1, Z_1, U_2, \tilde{U}_1, \tilde{X}_1,\tilde{Z}_1, \tilde{U}_2)\big)} &\leq & D_1\\
				\E{ d_2\big( S_2,\  \tilde{\phi}_{2,2}^*( V_1, X_2, Z_2, U_1, \tilde{U}_2, \tilde{X}_2,\tilde{Z}_2,\tilde{U}_2)\big)}&\leq & D_2
			\end{IEEEeqnarray}
		\end{subequations}	
		for some choice of  pmf $P_{U_1U_2X_1X_2Z_1Z_2}$ and functions $f_1$ and $f_2$ satisfying the stationarity condition \eqref{eq:stationarity} and $V_1,V_2$ satisfying the Markov chains \eqref{eq:Markov1}. 
	\end{theorem}

	Similarly to Shannon's inner bound, we can obtain the following corollary by setting $X_k=\tilde{U}_k$. 
	\begin{corollary}[Inner Bound via Non-Adaptive Coding]\label{cor:NonAdaptive}
		Any nonnegative rate-distortion quadruple $(R_1,R_2, D_1,D_2)$ is achievable if it  satisfies  the following two rate-constraints  
		\begin{subequations}\label{eq:HanCompression}
			\begin{IEEEeqnarray}{rCl} 
				R_1 & \leq & I(X_1;  X_2,Z_2) - I(V_1 ;  X_1, Z_1 | X_{2}, Z_2) \\
				R_2 & \leq & I(X_2; X_1,Z_1)- I(V_2 ;  X_2, Z_2 | X_{1}, Z_1),
			\end{IEEEeqnarray}
			and the two distortion constraints
			\begin{IEEEeqnarray}{rCl} 
				\E{ d_1\big( S_1,\  \tilde{\phi}_{2,1}^*( V_2, X_1, X_2, Z_1)\big) } &\leq & D_1\\
				\E{ d_2\big( S_2,\  \tilde{\phi}_{2,2}^*( V_1,X_1, X_2, Z_2)\big) }&\leq & D_2
			\end{IEEEeqnarray}
		\end{subequations}	
		for some choice of  pmfs $P_{X_1}$, $P_{X_2}$, $P_{V_1|X_1,Z_1}$, and $P_{V_2|X_2,Z_2}$. 
	\end{corollary}
	As the following example shows, above corollary achieves the fundamental rate-distortion tradeoff for some channels. 
	\begin{example}
		Consider the following state-dependent two-way channel 
		\begin{subequations}\label{eq:HanCompression}
			\begin{IEEEeqnarray}{rCl} 
				Z_1 & = &X_1\oplus X_2\oplus S_2 \qquad \textnormal{and} \qquad 
				Z_2   =   X_1\oplus X_2\oplus S_1,
			\end{IEEEeqnarray}
		\end{subequations}
		where inputs, outputs, and states are binary and $S_1$ and $S_2$ are independent Bernoulli-$p_1$ and $p_2$ random variables, for $p_1,p_2\in[0,1/2]$. Notice that Terminal 1's outputs depend on the state desired at Terminal 2 and Terminal 2's outputs on the state desired at Terminal 1, which calls for collaborative sensing.
		
		%
		Whenever $D_{\bar{k}}<p_{\bar{k}}$, we chooose
		\begin{IEEEeqnarray}{rCl} 
			V_k&= &Z_k \oplus X_k \oplus B_k=X_{\bar{k}}\oplus S_{\bar{k}}\oplus B_k
		\end{IEEEeqnarray}
		where $B_k$ is an independent Bernoulli-$D_k$ random variable. If $D_k \geq p_k$, choose $V_k$ a constant. 
		Inputs  $X_1$ and $X_2$ are chosen independent Bernoulli-$1/2$, i.e., capacity-achieving on  channels with Bernoulli-noses.
		When $D_{\bar{k}}<p_{\bar{k}}$, the optimal  symbo-by-symbol state-estimator is
		\begin{equation}
			\tilde{\phi}_{2,{\bar{k}}}^*( v_k, x_1,x_2, z_{\bar{k}}) = v_k \oplus x_{\bar{k}}
		\end{equation}
		and otherwise it is the constant estimator $\tilde{\phi}_{2,{\bar{k}}}^*( v_k, x_1,x_2, z_{\bar{k}}) =0$. 
		
		For the described  choice of random variables,  Corollary \ref{cor:NonAdaptive}  evaluates to the set of rate-distortion tuples $(R_1,R_2,D_1,D_2)$ satisfying
		\begin{IEEEeqnarray}{rCl} \label{eq:reg}
			R_k & \leq &1- H_{\text{b}}(p_k) -  \max\{0,\ H_{\text{b}}(p_{\bar{k}})-H_{\text{b}}(D_{\bar{k}}) \}, \qquad k\in\{1,2\},
		\end{IEEEeqnarray}
		and achieves the fundamental rate-distortion region as we show through a converse in Appendix~\ref{app:converse}.
		The region in \eqref{eq:reg} is concave (because the rate-distortion function $ \max\{0,\ H_{\text{b}}(p_{\bar{k}})-H_{\text{b}}(D_{\bar{k}}) \}$ is convex), and thus improves over classic time- and resource-sharing schemes. It also improves over a similar ISAC scheme without collaborative sensing where the compression codewords $V_1$ and $V_2$ are set to constants. In this latter case, only rate-distortion tuples are possible that satisfy $D_k\geq p_k$, for $k\in\{1,2\}$. 
	\end{example}
	
	\textcolor{black}{
		\begin{remark}
			For certain channels and state-distributions Theorem~\ref{thm:Sep} can be improved with the idea of coded time-sharing. The same applies for Theorem~\ref{thm:JSCC} in the next-following section.
		\end{remark}
	}
	\subsection{Collaborative Sensing and JSCC Scheme}
	
	In this scheme, we fully integrate the compression into the communication scheme, in a similar way that  hybrid coding \cite{Minero2015Hybrid} uses a single codeword for compression and channel coding in source-channel coding applications. 
	
	\begin{figure}[h!]
		\centering
		\includegraphics[scale=0.33]{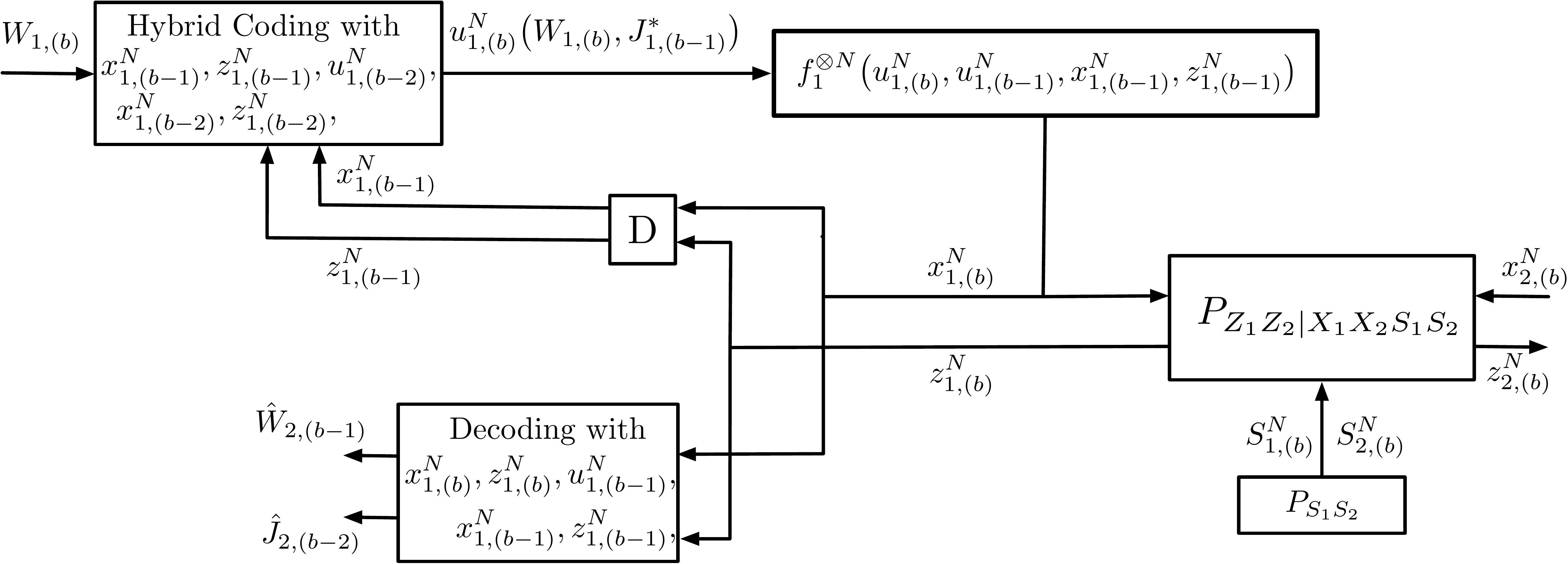}
		\caption{A ISAC scheme integrating collaborative sensing for D2D into Han's two-way coding scheme by means of hybrid coding. A single codeword is used both for compression and for channel coding.}
		\label{fig:TW3}
		\hspace{-10cm}
	\end{figure}

	Encoding and decoding in block $b$ of the new scheme are  depicted in Figure~\ref{fig:TW3}. The main difference compared to the scheme in the previous subsection is that here the  block-$b$ codeword $u_{1,(b)}^N$ is  \emph{correlated} with the inputs and outputs in the previous block $(b-1)$.\footnote{In the previous scheme, the compression codeword $v_{1,(b)}^N$ was correlated with the block-$(b-1)$ signals but not the channel coding codeword \textcolor{black}{$u^N_{1,(b)}$}. Now the codeword $u_{1,(b)}^N$ acts  both as a compression codeword  and as a channel coding codeword.} This correlation introduces additional dependence between blocks, which was previously missing because of the independence of the compression codewords and the codewords used for channel coding in the next block. To still obtain a stationary distribution on the codewords and channel inputs/outputs, which then allows for a single-letter characterization of the performance of the scheme, one has to choose a joint pmf $P_{U_1'U_2'Z_1Z_2X_1X_2U_1U_2}$, conditional pmfs $P_{U_1'|X_1Z_1\tilde{U}_1\tilde{X}_1\tilde{Z}_1}$ and $P_{U_2'|X_2Z_2\tilde{U}_2\tilde{X}_2\tilde{Z}_2}$ as well as  functions $f_1$ and $f_2$ on appropriate domains satisfying the new stationarity condition
	\begin{IEEEeqnarray}{rCl} 
		\lefteqn{P_{U_1'U_2'Z_1Z_2X_1X_2U_1U_2}(u_1',u_2',z_1,z_2,x_1,x_2)}\quad \nonumber \\
		&=& \sum_{\tilde{u}_1,\tilde{u}_2, \tilde{x}_1,\tilde{x}_2,\tilde{z}_1,\tilde{z}_2} P_{U_1'|X_1Z_1\tilde{U}_1\tilde{X}_1\tilde{Z}_1} (u_1'|u_1,x_1,z_1,\tilde{u}_1,\tilde{x}_1,\tilde{z}_1) P_{U_2'|X_2Z_2\tilde{U}_2\tilde{X}_2\tilde{Z}_2} (u_2'|x_2,z_2,u_2,\tilde{u}_2,\tilde{x}_2,\tilde{z}_2)  	\nonumber \\
		&&  \hspace{2cm}\cdot P_{Z_1Z_2|X_1X_2}(z_1,z_2|x_1,x_2)
		\mathbbm{1}\{x_1=f_1(u_1,\tilde{u}_1,\tilde{x}_1,\tilde{z}_1)\} \mathbbm{1}\{x_2=f_2(u_2,\tilde{u}_2,\tilde{x}_2,\tilde{z}_2)\}    \nonumber \\
		&& \hspace{4cm}\cdot P_{U_1'U_2'Z_1{Z}_2{X}_1X_2{U}_1{U}_2}(u_1,u_2, \tilde{z}_1,\tilde{z}_2,\tilde{x}_1,\tilde{x}_2,\tilde{u}_1,\tilde{u}_2),\label{eq:stationarity2}
	\end{IEEEeqnarray}
	In the following, all mentioned conditional and marginal pmfs are with respect  to the joint pmf $P_{U_1'U_2'Z_1Z_2X_1X_2U_1U_2\tilde{U}_1\tilde{U}_2\tilde{W}_1\tilde{W}_2\tilde{V}_1\tilde{V}_2}$ indicated by the summand in \eqref{eq:stationarity2}. 
	
	We next explain the code construction, encodings and decodings. 

	For each $k\in\{1,2\}$, for each block $b\in\{1,\ldots, B+1\}$, and  each message $m_k\in[2^{N\bar{R}_k}]$, choose a subcodebook $\{u_{k,(b)}^N(m_k,j) \colon j\in [2^{N R_k'}]\}$ by picking all entries i.i.d. $P_{U_k'}$. Terminal $k$ then picks the codeword $u_{k,(b)}^N(W_{k,(b)},j)$ so that the following joint-typicality check is satisfied for some fixed $\epsilon>0$:
	\begin{IEEEeqnarray}{rCl} 
		\big( u_{k,(b)}^N(W_{k,(b)}, j ),  x_{k,(b-1)}^N, \, z_{k,(b-1)}^N, \,  u_{k,(b-2)}^N, \, x_{k,(b-2)}^N, \, z_{k,(b-2)}^N \big)  \in \mathcal{T}_{\epsilon}^{N}\Big( P_{U_k' X_k Z_k \tilde{U}_k  \tilde{X}_k\tilde{Z}_k }\Big),  \IEEEeqnarraynumspace
	\end{IEEEeqnarray}
	and sets $J_{k,(b-1)}^*=j$. 
	By standard arguments, such an index $j$  exists with probability tending to 1 as $N\to \infty$ if
	\begin{IEEEeqnarray}{rCl}
		{R}_k '&\geq & I(U_k'; X_k, Z_k, \tilde{U}_k, \tilde{X}_k, \tilde{Z}_k), \quad k\in\{1,2\}. 
	\end{IEEEeqnarray}
	Terminal $k$ then sends the block-$b$ input sequence 
	\begin{equation}
		X_{k,(b)}^N = f_k^{\otimes N} \left( u_{k,(b)}^N\big(W_{k,(b)} J_{k,(b-1)}^*\big), u_{k,(b-1)}^N, x_{k,(b-1)}^n, z_{k,(b-1)}^N \right).
	\end{equation}
	
	Decoding is again performed  using a joint-typicality decoder. At the end of block $b$, Terminal~$k$ looks for indices $\hat{w}_{\bar{k}}$ and $\hat{j}_{\bar{k}}$ satisfying the two typicality checks
	\begin{IEEEeqnarray}{rCl} 
		\big( u_{\bar{k},(b-1)}^N(\hat{w}_{\bar{k}}, \hat{j}_{\bar{k}} ),  x_{k,(b)}^N, \, z_{k,(b)}^N,\, u_{k,(b-1)}^N  \, x_{k,(b-1)}^N, \, z_{k,(b-1)}^N \big)  \in \mathcal{T}_{\epsilon}^{N}\Big( P_{\tilde{U}_{\bar{k}} X_k Z_k \tilde{U}_k  \tilde{X}_k\tilde{Z}_k }\Big)
	\end{IEEEeqnarray}
	and 
	\begin{IEEEeqnarray}{rCl} 
		\big( u_{\bar{k},(b-1)}^N(\hat{w}_{\bar{k}}, \hat{j}_{\bar{k}} ),  x_{k,(b-2)}^N, \, z_{k,(b-2)}^N, \,  u_{k,(b-3)}^n, \, x_{k,(b-3)}^N, \, z_{k,(b-3)}^N \big)  \in \mathcal{T}_{\epsilon}^{N}\left( P_{U_{\bar{k}} X_k Z_k \tilde{U}_k  \tilde{X}_k\tilde{Z}_k }\right). \label{eq:WZJT}
	\end{IEEEeqnarray}
	If a unique pair of such element exists, set  $\hat{W}_{\bar{k},(b-1)}=w_{\bar{k}}$ and 
	$\hat{u}_{\bar{k},(b-1)}^N\eqdef u_{\bar{k},(b-1)}^N(\hat{w}_{\bar{k}}, \hat{j}_{\bar{k}} )$.
	Decoding is successful 
	with probability tending to 0 as $N\to \infty$ if 
	\begin{IEEEeqnarray}{rCl}
		\bar{R}_{\bar{k}}+{R}_{\bar{k}}'&\leq & I(\tilde{U}_{\bar{k}}; X_k, Z_k, \tilde{U}_k, \tilde{X}_k, \tilde{Z}_k) +  I(U_{\bar{k}}; X_k, Z_k, \tilde{U}_k, \tilde{X}_k, \tilde{Z}_k), \quad k\in\{1,2\}. 
	\end{IEEEeqnarray}
	
	State-estimation is  similar to \eqref{eq:state_estimate1}, but where Terminal $k$  replaces the compression codeword $v_{k,(b)}^N$  by the joint source-channel  codeword $u_{k,(b+1)}^N$  and similarly to hybrid coding also uses the inputs/outputs corresponding to the block where the codeword $u_{1,(b+1)}^N$ is sent, i.e., inputs and outputs in block $b+1$. Thus, Terminal $k$ computes its estimate of the block-$b$ state as:
	\begin{equation}\label{eq:state_estimate2}
		\hat{s}^N_{k,(b)} = \phi_{2,k}^{*\otimes N} \big(   \hat{u}_{\bar{k},(b+1)}^N,x_{k,(b+1)}^N,  z_{k,(b+1)}^N,   \hat{u}_{\bar{k},(b)}^N, x_{k,(b)}^N,  z_{k,(b)}^N, u_{k,(b-1)}^N, x_{k,(b-1)}^N,  z_{k,(b-1)}^N,  \hat{u}_{\bar{k},(b-1)}^N\big),
	\end{equation}
	where 
	\begin{IEEEeqnarray}{rCl}
		{\phi}_{2,k}^*(u_{\bar{k}}', x_k', z_k',u_{\bar{k}},  x_k,z_k,  \tilde{u}_k, \tilde{x}_{k}, \tilde{z}_k, \tilde{u}_{\bar{k}}):= 
		\textnormal{arg}\min_{s_k'\in \hat{\mathcal{S}_k}} \sum_{s_k\in \mathcal{S}_k}  P_{S_k|X_kZ_kU_{\bar{k}}}(s_k|x_k,z_k,u_{\bar{k}})\;  d_k(s_k,  s_k').\nonumber\\
	\end{IEEEeqnarray}
	By standard  arguments and because of the stationarity condition in \eqref{eq:stationarity2}  
	the probability of violating the distortion constraints tends to 0 as $N\to \infty$ if 
	\begin{subequations}\label{eq:dist_constraintsJSCC}
		\begin{IEEEeqnarray}{rCl} 
			\E{ d_k\big( S_k,\  \phi_{2,k}^*( U_{\bar{k}}', X_k', Z_k',U_{\bar{k}},X_k, Z_k, \tilde{U}_k, \tilde{X}_k,\tilde{Z}_k,\tilde{U}_{\bar{k}})\big)} &\leq & D_k, \qquad k\in\{1,2\},
		\end{IEEEeqnarray}
	\end{subequations}
	where $X_1'=f_1(U_1', U_1, X_1, Z_1)$ and $X_2'=f_2(U_2', U_2, X_2, Z_2)$ and the outputs $Z_1'$ and $Z_2'$ are obtained from $X_1'$ and $X_2'$ via the channel transition law $P_{Z_1Z_2|X_1X_2}$.

	From above considerations and by eliminating the dummy rates ${R}_1'$ and ${R}_2'$, we obtain the following theorem.

	\begin{theorem}[Inner Bound via Joint Source-Channel Coding]\label{thm:JSCC}
		Any nonnegative rate-distortion quadruple $(R_1,R_2, D_1,D_2)$ is achievable if it  satisfies  the following two rate-constraints  
		\begin{IEEEeqnarray}{rCl} 
			{R}_{{k}}&\leq & I(\tilde{U}_{{k}}; X_{\bar{k}}, Z_{\bar{k}}, \tilde{U}_{\bar{k}}, \tilde{X}_{\bar{k}}, \tilde{Z}_{\bar{k}}) 
			-  I(U_{{k}}; X_k, Z_k, \tilde{U}_k, \tilde{X}_k, \tilde{Z}_k| X_{\bar{k}}, Z_k, \tilde{U}_{\bar{k}}, \tilde{X}_{\bar{k}}, \tilde{Z}_{\bar{k}}), \quad k\in\{1,2\} \label{eq:JSCC}\IEEEeqnarraynumspace
		\end{IEEEeqnarray}
		and the two distortion constraints  in \eqref{eq:dist_constraintsJSCC}
		for some choice of  pmf $P_{U_1'U_2'Z_1Z_2X_1X_2U_1U_2}$ and functions $f_1$ and $f_2$ satisfying the stationarity condition \eqref{eq:stationarity}. 
	\end{theorem}
	
	
	\begin{remark}We notice that the described compression technique does not use binning as in Wyner-Ziv coding\cite{WynerZiv76}. Instead,  
		decoder side-information is taken into account via the joint  typicality check in \eqref{eq:WZJT}. 
	\end{remark}	
	
	\begin{remark} \label{rem:separate}
		For the choice $U_k'=(U_k'', V_k)$ with $U_k''\sim P_{U_k}$ independent of all other random variables and $V_{1}$ and $V_2$ satisfying the Markov chains in \eqref{eq:Markov1}, the inner bound in Theorem~\ref{thm:JSCC} achieved by our joint source-channel coding scheme specializes to the inner bound Theorem~\ref{thm:Sep} achieved by  separate source-channel coding. For above choice of auxiliary random variables, the reconstruction functions $g_1$ and $g_2$ can restrict  their first arguments only to the $V_1$- and $V_2$-components  without loss in performance.
	\end{remark}

	\section{Summary and Outlook}	
	We considered integrated sensing and communication (ISAC) over multi-access channels (MAC) and device-to-device (D2D) communication, where different terminals help each other to improve sensing.
	We reviewed related communication schemes and proposed adaptations that fully integrate the collaborative sensing into information-theoretic data communication schemes. For D2D communication, we also proposed a joint source-channel coding (JSCC) scheme to integrate compression and coding into a single codeword as in hybrid coding.  Through examples, we  demonstrated the advantages of our collaborative sensing ISAC schemes compared to  non-collaborative ISAC schemes  with respect to the achieved rate-distortion regions.
	Various interesting future research directions arise. As already mentioned, the JSCC scheme proposed for ISAC D2D communication could be integrated into our ISAC MAC scheme. Another interesting research direction for the MAC scheme is to include state-estimation at the Rx. In this respect, it would be interesting to include an additional superposition compression layer to generate compression information that is only decoded by the Rx but not the other Tx. For D2D communication an interesting extension would be to consider specific channel models and to replace Han's result by two-way communication schemes that are tailored to these specific channels. 
	
	\section*{Acknowledgment}
	The authors would like to thank Mari Kobayashi for helpful discussions. 
	\appendices%
	\section{Converse to Example~3}\label{app:converse}
	By the independence of the messages and Fano's Inequality, we obtain for some function $\epsilon_n$ that vanishes as $n\to \infty$,
	\begin{IEEEeqnarray}{rCl}
		R_k & \leq  & \frac{1}{n} I(W_k; Y_{\bar{k}}^n|W_{\bar{k}}) + \epsilon_n \\
		& =  &  \frac{1}{n} I(W_k S_{\bar{k}}^n; Y_{\bar{k}}^n|W_{\bar{k}})  -   \frac{1}{n} I( S_{\bar{k}}^n; Y_{\bar{k}}^n| W_kW_{\bar{k}}) + \epsilon_n \\
		&= & \frac{1}{n} \left[ \sum_{i=1}^n I(Y_{\bar{k},i} ; W_k S_{\bar{k}}^n | Y_{\bar{k}}^{i-1} W_{\bar{k}} )-I( S_{\bar{k},i}; Y_{\bar{k}}^n| W_k  W_{\bar{k}}S_{\bar{k}}^{i-1})  \right]   +\epsilon_n \\
		& \stackrel{(a)}{\leq}  & \frac{1}{n} \left[ \sum_{i=1}^n I(Y_{\bar{k},i} ; X_{k,i}W_k S_{\bar{k}}^n|X_{\bar{k},i} Y_{\bar{k}}^{i-1}W_{\bar{k}} )-I( S_{\bar{k},i}; Y_{\bar{k}}^n W_k W_{\bar{k}}S_{\bar{k}}^{i-1})  \right]   +\epsilon_n \\
		&\stackrel{(b)}{\leq}   & \frac{1}{n} \left[ \sum_{i=1}^n I(Y_{\bar{k},i} ; X_{{k},i})-I( S_{\bar{k},i}; \hat{S}_{\bar{k},i})  \right]   +\epsilon_n \\
		&\stackrel{(c)}{\leq} & nC_{{k}} - R_{\bar{k}}(D_{\bar{k}})+\epsilon_n,
	\end{IEEEeqnarray}
	where $C_{{k}}$ denotes the capacity of the point-to-point channel from $X_k$ to $X_{k}+S_{k}$  and $R_{\bar{k}}(\cdot)$ denotes the rate-distortion function of source $S_{\bar{k}}$. In our example, $C_{\bar{k}}=1-H_{\textnormal{b}}(p_k)$ and $R_{\bar{k}}(D_{\bar{k}})=[H_{\textnormal{b}}(p_k)-H_{\textnormal{b}}(D_k)]^+$. Justification for above inequalities are as follows: 
	$(a)$ holds because conditioning \textcolor{black}{cannot increase} entropy, because $X_{\bar{k},i}$ is a function of $W_{\bar{k}}$ and $Y_{\bar{k}}^{i-1}$, and  by the i.i.d.ness of the source sequence $S_{{k}}^n$;
	$(b)$ holds because of the Markov chain $Y_{\bar{k},i} \to( X_{k,i} , X_{\bar{k},i})\to (W_k ,W_{\bar{k}},S_{\bar{k}}^n ,Y_{\bar{k}}^{i-1})$ and because $\hat{S}_{\bar{k},i}$ is a function of $Y_{\bar{k}}^n$ and again because  conditioning \textcolor{black}{cannot increase} entropy;
	$(c)$ holds  by the definition of the rate-distortion function $R_{\bar{k}}(\cdot)$  and because  $R_{\bar{k}}(\cdot)$ is convex and monotonic. 		
	
	%
	%

	\lo{
		\section{Proof of Theorem~\ref{Th:achievability:MAC}}\label{app:analysis}
		To derive an upper bound  on the average error probability (averaged over the random code construction and the state and channel realizations), we enlarge the error event to the event that for some $k=1,2$ and $b=1,\ldots, B$:
		\begin{IEEEeqnarray}{rCl}
			\hat{W}_{k,c,(b)} \neq W_{k,c,(b)}\quad \text{or} \quad 
			\hat{W}_{k,p,(b)}\neq W_{k,p,(b)}\quad 	
			\text{or} \quad 	\hat{W}_{k,c,(b)}^{(\bar k)} \neq W_{k,c,(b)}	
		\end{IEEEeqnarray}
		or 
		\begin{IEEEeqnarray}{rCl}
			J_{k,(b)}^*= -1 \quad \text{or} \quad 	\hat{J}_{k,(b)}\neq J^*_{k,(b)} \quad \text{or} \quad \hat{J}_{k,(b)}^{(\bar{k})} \neq J^*_{k,(b)}.\IEEEeqnarraynumspace
		\end{IEEEeqnarray}
		For ease of notation, we define the block-$b$ Tx-error events  for $k=1,2$ and $b=1,\ldots, B$:
		\begin{IEEEeqnarray}{rCl}
			\mathcal{E}_{\textnormal{Tx},k,(b)}:=\Big \{  \hat{W}^{(k)}_{\bar{k},c,(b)}\neq W_{\bar{k},c,(b)} 	\;\text{ or }\;
			\hat{J}^{(k)}_{\bar{k},(b-1)}\neq J^*_{\bar{k},(b-1)} 
			\hspace{0cm}	\;\text{ or }\;J_{k,b}^*= -1 \Big\},\IEEEeqnarraynumspace
		\end{IEEEeqnarray}
		and
		\begin{IEEEeqnarray}{rCl}
			\mathcal{E}_{\textnormal{Tx},k,(B+1)}:=\left\{ 
			\hat{J}^{(k)}_{\bar{k},(B)}\neq J^*_{\bar{k},(B)}\right \}, \qquad k \in\{1,2\}.
		\end{IEEEeqnarray}
		
		Define also the Rx-error events for $k=1,2$ and block $b=1,\ldots, B+1$:
		\begin{IEEEeqnarray}{rCl}
			\mathcal{E}_{\textnormal{Rx},(b)}:=\Big\{  \hat{W}_{k,c,(b-1)}\neq W_{k,c,(b-1)} 	\;\text{ or }\; \hat{W}_{k,p,(b)}\neq W_{k,p,(b)} 
			\hspace{0cm}\;\text{ or }\; 
			\hat{J}_{k,(b-1)}\neq J^*_{k,(b-1)}\colon \; k=1,2\Big\}.\IEEEeqnarraynumspace
		\end{IEEEeqnarray}
		
		By the union bound and basic probability, we  find:
		\begin{IEEEeqnarray}{rCl}\label{error_P_def}
			\Pr\left( \hat{W}_1 \neq W_1 \;\text{ or }\; \hat{W}_2 \neq W_2\right)
			\leq	
			\sum_{b=1}^{B+1} 	\Pr\left( \mathcal{E}_{\textnormal{Rx},(b)} \Bigg|   \bigcup_{b'=1}^{B+1}\left\{ \bar{ \mathcal{E}}_{\textnormal{Tx},1,(b')}, \; \bar{ \mathcal{E}}_{\textnormal{Tx},2,(b')}\right\} \right)
			&&
			\nonumber \qquad \\
			&&\hspace{-6cm}+	\sum_{b=1}^{B+1}
			\Pr\left( \mathcal{E}_{\textnormal{Tx},1,(b)}\Bigg| \bigcup_{b'=1}^{b-1} \left\{ \bar{ \mathcal{E}}_{\textnormal{Tx},1,(b')}, \; \bar{ \mathcal{E}}_{\textnormal{Tx},2,(b')} \right\}
			\right) 
			\nonumber \\
			&&\hspace{-5cm}
			+
			\sum_{b=1}^{B+1} 	\Pr\left( \mathcal{E}_{\textnormal{Tx},2,(b)}\Bigg| \bigcup_{b'=1}^{b-1} \left\{  \bar{ \mathcal{E}}_{\textnormal{Tx},1,(b')}, \; \bar{ \mathcal{E}}_{\textnormal{Tx},2,(b')} \right\}
			\right) 
			.\IEEEeqnarraynumspace
			\label{analysis:part1}
		\end{IEEEeqnarray}
		
		We   analyze the three sums separately. The first sum  is related to Tx~1's error event, the second sum  to Tx~2's error event, and the third sum to the Rx's error event.
		\subsubsection{Analysis of Tx~1's error event} To simplify notations, we define for each block $b\in\{2,\ldots, B+1\}$ and each triple of indices $(j_1^*, \hat{w}_2, \hat{j}_2)$ 
		the event $\mathcal{F}_{\textnormal{Tx1},(b)}(j_1^*, \hat{w}_2, \hat{j}_2)$ that the following two conditions  \eqref{typ1:enc_1b} and \eqref{typ2:enc_1b}  (only Condition  \eqref{typ1:enc_1b} for \mw{$b=1$}) hold: 
		\begin{IEEEeqnarray}{rCl}\label{typ1:enc_1b}
			&&\hspace{-1cm}	 \Bigg(
			u_{0,(b)}^N\Big(W_{1,c,(b-1)}, \hat{W}_{2,c,(b-1)}^{(1)}\Big),
			\;
			u^N_{1,(b)}\Big(W_{1,c,(b)}, J^*_{1,(b-1)}\; \Big| \;
			W_{1,c,(b-1)}, {W}_{2,c,(b-1)}
			\Big)  \qquad 
			\nonumber
			\\
			&&\quad
			u_{2,(b)}^N\Big( \hat{w}_{2},\hat{j}_{2} \;
			\Big| \;W_{1,c,(b-1)}, {W}_{2,c,(b-1)}
			\Big), \;  x_{1,(b)}^N\Big(W_{1,p,(b)}\;\Big| \;
			W_{1,c,(b)}, J^*_{1,(b-1)}, W_{1,c,(b-1)}, {W}_{2,c,(b-1)}\Big), 
			\nonumber
			\\
			&&\hspace{0.7cm} \quad
			v_{1,(b)}^N\Big(j^*_{1}\;\Big|\; {J^*_{1,(b-1)}}, 
			W_{1,c,(b)}, \hat{w}_{2}, \hat{j}_{2} ,
			W_{1,c,(b-1)}, {W}_{2,c,(b-1)}
			\Big), \; 
			Z^N_{1,(b)}
			\Bigg) 
			\nonumber\\
			&&\hspace{11cm}
			\in \mathcal{T}^N_{\epsilon}(P_{U_0U_1U_2X_1 V_1 Z_1})
		\end{IEEEeqnarray}
		and if $b >1$
		\begin{IEEEeqnarray}{rCl}\label{typ2:enc_1b}
			&&\hspace{-1cm}\Bigg(
			u_{0,(b-1)}^N\Big(
			W_{1,c,(b-2)}, 
			{W}_{2,c,(b-2})\Big), \; 
			\nonumber	\\
			&& \hspace{-0cm}\quad u_{1,(b-1)}^N\Big(
			W_{1,c,(b-1)},
			{J}_{1,(b-1)}^*
			\; \Big| \; 
			W_{1,c,(b-2)}, 
			{W}_{2,c,(b-2)}
			\Big) 
			\nonumber	\\
			&&\hspace{1cm}\quad u_{2,(b-1)}^N\Big( {W}_{2,c,(b-1)}, {J}_{2,(b-2)}
			\Big| \; W_{1,c,(b-2)}, {W}_{2,c,(b-2)}
			\Big), \; 
			\nonumber	\\
			&&\hspace{2cm}\quad x_{1,(b-1)}^N\Big(W_{1,p,(b-1)}\; \Big| \; W_{1,c,(b-1)}, J^*_{1,(b-2)}, 
			W_{1,c,(b-2)}, {W}_{2,c,(b-2)}\Big), 
			\nonumber		\\
			&&\hspace{3cm}\quad v_{2,(b-1)}^N\Big(
			\hat{j}_{2}
			\; \Big| \;			
			W_{1,c,(b-1)}, 
			J^*_{1,(b-2)}
			,
			{W}_{2,c,(b-1)},{J}_{2,(b-2)}^*,
			W_{1,c,(b-2)}, {W}_{2,c,(b-2)}\Big), \;  
			\nonumber	\\
			&&\hspace{9cm}\quad 	Z^N_{1,(b-1)} \Bigg)
			\in \mathcal{T}^N_{\epsilon}(P_{U_0U_1U_2X_1 V_2 Z_1}).
		\end{IEEEeqnarray}
		Notice that compared to \eqref{typ1:enc_1} and \eqref{typ2:enc_1}, here we replaced the triple $(\hat{W}_{2,c,(b-2)}^{(1)},\hat{W}_{2,c,(b-1)}^{(1)}, \hat{J}_{2,(b-2)}^{(1)})$ by their correct values $W_{2,c,(b-2)}, W_{2,c,(b-1)}, J^*_{2,(b-2)})$.
		Similarly, define the event $\mathcal{F}_{\textnormal{Tx1},(B+1)}(\hat{j}_2)$ as the event that the following two conditions are  satisfied: 
		\begin{IEEEeqnarray}{rCl}\label{typ1:enc_1d}
			&&\hspace{-1cm} \Bigg(
			u_{0,(B+1)}^N\Big(W_{1,c,(B)}, {W}_{2,c,(B)}\Big),
			\;
			\nonumber		\\
			&&\hspace{-0cm}\quad 	u^N_{1,(B+1)}\Big(1, J^*_{1,(B)} \; \Big| \;
			W_{1,c,(B)}, {W}_{2,c,(B)}\Big),\;  
			u_{2,(B+1)}^N\Big(1,\hat{j}_{2} \;
			\Big| \;W_{1,c,(B)}, {W}_{2,c,(B)}
			\Big), \; 
			\nonumber		\\
			&&\hspace{1cm}\quad 
			x_{1,(B+1)}^N\Big(1\;\Big| \;
			W_{1,c,(B+1)}, J^*_{1,(B)}, W_{1,c,(B)}, {W}_{2,c,(B)}\Big), 
			\; Z^N_{1,(B+1)}
			\Bigg) \in \mathcal{T}^N_{\epsilon}(P_{U_0U_1U_2X_1 Z_1})
		\end{IEEEeqnarray}
		and
		\begin{IEEEeqnarray}{rCl}\label{typ2:enc_1d}
			&&\hspace{-1cm}{ \Bigg(u_{0,(B)}^N\Big(
				W_{1,c,(B-1)}, 
				{W}_{2,c,(B-1)} \Big), \; } \quad
			\nonumber \\
			&&\hspace{-0cm} 	u_{1,(B)}^N\Big(
			W_{1,c,(B)},
			{J}_{1,(B)}^*
			\; \Big| \; 
			W_{1,c,(B-1)}, 
			{W}_{2,c,(B-1)}
			\Big) ,
			\nonumber	\qquad	\\
			&&\hspace{1cm}u_{2,(B)}^N\Big( {W}_{2,c,(B)}^{(1)}, {J}_{2,(B-1)}^{(1)}\; \Big| \; W_{1,c,(B-1)}, \hat{W}_{2,c,(B-1)}^{(1)}
			\Big), \; 
			\nonumber \\
			&&\hspace{2cm} x_{1,(B)}^N\Big(W_{1,p,(B)}\; \Big| \; W_{1,c,(B)}, J^*_{1,(B-1)}, 	W_{1,c,(B-1)}, \hat{W}_{2,c,(B-1)}^{(1)}\Big), 
			\nonumber		\\
			&&\hspace{3cm} 
			v_{2,(B)}^N\Big(
			\hat{j}_{2}
			\; \Big| \; 
			W_{1,c,(B)}, 
			J^*_{1,(B-1)}
			,
			\hat{W}_{2,c,(B)}^{(1)},{J}_{2,(B-1)}^{(1)}, W_{1,c,(B-1)}, {W}_{2,c,(B-1)}\Big), \; 
			\nonumber\\
			&&\hspace{9cm}Z^N_{1,(B)} \bigg)
			\in \mathcal{T}^N_{\epsilon}(P_{U_0U_1U_2X_1 V_2 Z_1})
		\end{IEEEeqnarray}

		We continue by noticing that  event $\bigcup_{b'=1}^{b-1} \left\{  \bar{\mathcal{E}}_{\textnormal{Tx},1,(b')}, \; \bar{\mathcal{E}}_{\textnormal{Tx},2,(b')} \right\}$ implies  that for all $b'=1, \ldots, b-1$, $k =1,2$:
		\begin{IEEEeqnarray}{rCl}
			\hat{W}^{(k)}_{\bar{k},c,(b')}& =&W_{\bar{k},c,(b')}\\
			J_{k,(b')}^*&\neq&- 1\\
			\hat{J}^{(\bar{k})}_{k,(b'-1)}&=& J_{{k},(b'-1)}^*.
		\end{IEEEeqnarray}
		Moreover, for any block $b=1,\ldots, B+1$,  event  $ \bar{\mathcal{E}}_{\textnormal{Tx},1,(b)}$ is implied by the  event that $\mathcal{F}_{\textnormal{Tx1},(b)}(j_1^*, \hat{w}_{2},\hat{j}_2)$ \emph{is  not} satisfied for any tuple $(j_1^*, \hat{w}_{2},\hat{j}_2)$ with  $(\hat{w}_{2},\hat{j}_2)=( {W}_{2,c,(b)}, J_{2,(b-1)}^*)$ or it \emph{is} satisfied for some triple $(j_1^*, \hat{w}_{2},\hat{j}_2)$ with $( \hat{w}_{2},\hat{j}_2)\neq ({W}_{2,c,(b)}, J_{2,(b-1)}^*)$. 
		Thus,  the sequence of inequalities on top of the next page holds,
		\begin{figure*}
			\begin{subequations}
				\begin{IEEEeqnarray}{rCl}
					\lefteqn{\Pr\left( \mathcal{E}_{\textnormal{Tx},1,(b)} \; \Bigg|  \; \bigcup_{b'=1}^{b-1} \left\{ \bar{ \mathcal{E}}_{\textnormal{Tx},1,(b')}, \; \bar{ \mathcal{E}}_{\textnormal{Tx},2,(b')} \right\}
						\right) }  \nonumber \\
					& = &\Pr\Bigg( \Bigg(\bigcap_{j_1^*\in [2^{nR_{v,1}}] }\bar{\mathcal{F}}_{\textnormal{Tx1},(b)}(j_1^*, W_{2,c,(b)},J^*_{2,(b-1)})  \Bigg) \nonumber \\
					&&	 \qquad  \cup  \Bigg( \bigcup_{\substack{ (j_1^*, \hat{w}_{2},\hat{j}_2) \colon \\ ( \hat{w}_{2},\hat{j}_2)\neq ({W}_{2,c,(b)}, J_{2,(b-1)}^*) }}  \mathcal{F}_{\textnormal{Tx1},(b)}(j_1^*, \hat{w}_{2},\hat{j}_2)    \Bigg)\; \Bigg| \; \bigcup_{b'=1}^{b-1} \left\{ \bar{ \mathcal{E}}_{\textnormal{Tx},1,(b')}, \; \bar{ \mathcal{E}}_{\textnormal{Tx},2,(b')} \right\}
					\Bigg)  \nonumber \\\\
					& \leq  & \Pr\left( \bigcap_{j_1^*\in [2^{nR_{v,1}}] }\bar{\mathcal{F}}_{\textnormal{Tx1},(b)}(j_1^*, W_{2,c,(b)},J^*_{2,(b-1)})  \; \Bigg| \;\bigcup_{b'=1}^{b-1} \left\{ \bar{ \mathcal{E}}_{\textnormal{Tx},1,(b')}, \; \bar{ \mathcal{E}}_{\textnormal{Tx},2,(b')} \right\}
					\right) 
					\nonumber\\
					& & + \Pr\Bigg(  \bigcup_{\substack{ (j_1^*, \hat{w}_{2},\hat{j}_2) \colon \\ ( \hat{w}_{2},\hat{j}_2)\neq ({W}_{2,c,(b)}, J_{2,(b-1)}^*) }}  \mathcal{F}_{\textnormal{Tx1},(b)}(j_1^*, \hat{w}_{2},\hat{j}_2)  \; \Bigg| \; \bigcup_{b'=1}^{b-1} \left\{ \bar{ \mathcal{E}}_{\textnormal{Tx},1,(b')}, \; \bar{ \mathcal{E}}_{\textnormal{Tx},2,(b')} \right\}
					\Bigg) \\
					& \leq & 	 \Pr\left( \bigcap_{j_1^*\in [2^{nR_{v,1}}] }\bar{\mathcal{F}}_{\textnormal{Tx1},(b)}(j_1^*, W_{2,c,(b)},J^*_{2,(b-1)})  \; \Bigg| \;\bigcup_{b'=1}^{b-1} \left\{ \bar{ \mathcal{E}}_{\textnormal{Tx},1,(b')}, \; \bar{ \mathcal{E}}_{\textnormal{Tx},2,(b')} \right\}
					\right) 
					\nonumber\\
					& & + \sum_{\substack{ (j_1^*, \hat{w}_{2},\hat{j}_2) \colon \\ \hat{w}_{2} \neq W_{2,c,(b)}, \\ \hat{j}_2 \neq  J_{2,(b-1)}^*}} 
					\Pr\left(  \mathcal{F}_{\textnormal{Tx1},(b)}(j_1^*, \hat{w}_{2},\hat{j}_2)  \; \Bigg| \; \bigcup_{b'=1}^{b-1} \left\{ \bar{ \mathcal{E}}_{\textnormal{Tx},1,(b')}, \; \bar{ \mathcal{E}}_{\textnormal{Tx},2,(b')} \right\} 
					\right) 
					\nonumber\\
					& & +  \sum_{\substack{ (j_1^*,\hat{j}_2) \colon \\  \hat{j}_2 \neq  J_{2,(b-1)}^*}} 
					\Pr\left(  \mathcal{F}_{\textnormal{Tx1},(b)}(j_1^*,\mw{W_{2,c,(b)}},\hat{j}_2)  \; \Bigg| \; \bigcup_{b'=1}^{b-1} \left\{ \bar{ \mathcal{E}}_{\textnormal{Tx},1,(b')}, \; \bar{ \mathcal{E}}_{\textnormal{Tx},2,(b')} \right\} 
					\right) 	
					\nonumber\\
					& & +  \sum_{\substack{ (j_1^*, \hat{w}_{2}) \colon \\  \hat{w}_{2} \neq W_{2,c,(b)} }} 
					\Pr\left(  \mathcal{F}_{\textnormal{Tx1},(b)}(j_1^*, \hat{w}_{2},\mw{{J}_{2,(b-1)}^*})  \; \Bigg| \; \bigcup_{b'=1}^{b-1} \left\{ \bar{ \mathcal{E}}_{\textnormal{Tx},1,(b')}, \; \bar{ \mathcal{E}}_{\textnormal{Tx},2,(b')} \right\} 
					\right) ,
					\label{eq:last}
				\end{IEEEeqnarray}
			\end{subequations}
			\hrule
		\end{figure*}
		where the inequalities hold by the union bound.
		By the Covering Lemma \cite{cover2006elements}, the way we construct the codebooks and  the weak law of large numbers, and because we condition on event  $\bar{\mathcal{E}}_{\textnormal{Tx},2,(b-1)}$ implying $J_{2,b-1}^*\neq -1$, the first summand in \eqref{eq:last} tends to 0 as $N\to \infty$ if 
		\begin{equation}\label{eq:R1v}
			R_{1,v}> I(V_1;X_1Z_1\mid U_0U_1U_2).
		\end{equation}
		By the way we constructed the codebooks, and standard information-theoretic arguments \cite{ElGamal}, the  sum in the second  line of \eqref{eq:last} tends to 0 as $N \to \infty$, if 
		\begin{IEEEeqnarray}{rCl}\label{eq:R2cv}
			\mw{R_{1,v}+}	R_{2,v}+R_{2,c}&<& I(U_2V_1;Z_1X_1\mid U_0U_1)
			+I(V_2;Z_1X_1\mid U_0U_1U_2), \IEEEeqnarraynumspace
		\end{IEEEeqnarray}	
		the sum in the third line of \eqref{eq:last} tends to 0 as $N\to \infty$ if 
		\begin{IEEEeqnarray}{rCl}\label{eq:R2vlower}
			\mw{R_{1,v}+}R_{2,v} &<& I(U_2V_1;Z_1X_1\mid U_0U_1)
			+I(V_2;Z_1X_1\mid U_0U_1U_2),
		\end{IEEEeqnarray}	
		and	the sum in the fourth line of \eqref{eq:last}  tends to $0$ as $N\to \infty$ if
		\begin{IEEEeqnarray}{rCl}\label{eq:R2c}
			\mw{R_{1,v}+}	R_{2,c}&<& I(Z_1X_1;U_2V_1\mid U_0U_1).
		\end{IEEEeqnarray}
		Since Condition \eqref{eq:R2vlower} is obsolete in view of \eqref{eq:R2cv}, we conclude that for any finite $B$ the sum of the  probability of errors $\sum_{b=1}^{B+1} 
		\Pr\left( \mathcal{E}_{\textnormal{Tx},1,(b)}\big| \bigcup_{b'=1}^{b-1} \left\{ \bar{ \mathcal{E}}_{\textnormal{Tx},1,(b')}, \; \bar{ \mathcal{E}}_{\textnormal{Tx},2,(b')} \right\}
		\right)$ tends to $0$ as $N\to \infty$ if  Conditions  \eqref{eq:R1v}, \eqref{eq:R2cv}, and \eqref{eq:R2c} are satisfied.	 
		
		\subsubsection{Analysis of Tx~2's error event} 
		By similar arguments, one can also prove that for finite $B$ the sum of the  probability of errors $\sum_{b=1}^{B+1}
		\Pr\left( \mathcal{E}_{\textnormal{Tx},2,(b)}\big| \bigcup_{b'=1}^{b-1} \left\{ \bar{ \mathcal{E}}_{\textnormal{Tx},1,(b')}, \; \bar{ \mathcal{E}}_{\textnormal{Tx},2,(b')} \right\}
		\right)$ tends to $0$ as $N\to \infty$ if  Conditions  \eqref{eq:Rkv},   \eqref{eq:Rkc}, and \eqref{eq:RkvRkc}, are satisfied for $k=2$.
		
		\subsubsection{Analysis of Rx's error event} 
		For each block $b=2,\ldots, B$ and each tuple $(w_{1,c}, w_{2,c}, w_{1,p}, w_{2,p}, j_1, j_2)$ define  $\mathcal{F}_{\textnormal{Rx},(b)}(w_{1,c}, w_{2,c}, w_{1,p}, w_{2,p}, j_1, j_2)$  as the event 
		\begin{IEEEeqnarray}{rCl}\label{typ:dec_bb}
			&&	\Bigg(
			u^N_{0,(b)}(w_{1,c}, w_{2,c}),\;
			u^N_{1,(b)}\Big({W}_{1,c,(b)},j_{1}\; \Big | \; w_{1,c}, w_{2,c} \Big),\; u^N_{2,(b)}\Big({W}_{2,c,(b)}, j_{2}\; \Big | \; w_{1,c}, w_{2,c}\Big), \qquad \nonumber \\
			&&\hspace{1cm} \quad  x^N_{1,(b)}\Big(w_{1,p}\; \Big  | \; {W}_{1,c,(b)}, j_{1}, w_{1,c}, w_{2,c}\Big), \; 
			x^N_{2,(b)}\Big( w_{2,p}\; \Big | \; {W}_{2,c,(b)}, j_{2}, w_{1,c}, w_{2,c}\Big)
			\nonumber		\\
			&&\quad
			v_{1,(b)}^N\Big({J}_{1,(b)}\; \Big | \; 
			{W}_{1,c,(b)},  {W}_{2,c,(b)},
			w_{1,c},  j_1, w_{2,c}, j_2
			\Big),\;	v^N_{2,(b)}({J}_{2,(b)}\mid {W}_{1,c,(b)},  {W}_{2,c,(b)},w_{1,c}, j_1, w_{2,c} , j_2
			), \nonumber \\
			&& \hspace{9cm}
			\mw{ Y^N_{(b)}} \Bigg)	\mw{ \in \Tc_{2\epsilon}(P_{U_0U_1U_2X_1X_2 Y})}.
		\end{IEEEeqnarray}
		We continue by noticing that for $b=2, \ldots, B$ event  $\bar{\mathcal{E}}_{\textnormal{Rx},(b)}$ is equivalent to the event that $\mathcal{F}_{\textnormal{Rx},(b)}(w_{1,c}, w_{2,c}, w_{1,p}, w_{2,p}, j_1, j_2)$ \emph{is  not} satisfied for the tuple 
		$(w_{1,c}, w_{2,c}, w_{1,p}, w_{2,p}, j_1, j_2)
		=(W_{1,c,(b-1)}, 
		{W}_{2,c,(b-1)}, $
		$W_{1,p,(b)}, W_{2,p,(b)}, J_{1,(b-1)}^*, J_{2,(b-1)}^*)$ or it \emph{is} satisfied for some tuple  $(w_{1,c}, w_{2,c}, w_{1,p}, w_{2,p}, j_1, j_2) \neq 
		(W_{1,c,(b-1)}, {W}_{2,c,(b-1)}, $
		$W_{1,p,(b)}, W_{2,p,(b)}, J_{1,(b-1)}^*, J_{2,(b-1)}^*)$.  
		Similarly for events $\bar{\mathcal{E}}_{\textnormal{Rx},(1)}$  and $\bar{\mathcal{E}}_{\textnormal{Rx},(B+1)}$.
		Thus, for $b\in\{2,\ldots, B\}$, the sequence of (in)equalities \eqref{event_error_rec}  holds,
		\begin{subequations}\label{event_error_rec}
			\begin{IEEEeqnarray}{rCl}
				\lefteqn{\Pr\left( \mathcal{E}_{\textnormal{Rx},(b)} \; \Bigg|  \; \bigcup_{b'=1}^{B+1} \left\{ \bar{ \mathcal{E}}_{\textnormal{Tx},1,(b')}, \; \bar{ \mathcal{E}}_{\textnormal{Tx},2,(b')} \right\}
					\right) }  \nonumber \\
				& = &\Pr \Bigg(  \Bigg( \bigcup_{\substack{ (w_{1,c}, w_{2,c}, w_{1,p}, w_{2,p}, j_1, j_2)   \neq \\(W_{1,c,(b-1)}, {W}_{2,c,(b-1)}, W_{1,p,(b)}, W_{2,p,(b)}, J_{1,b-1}^*, J_{2,(b-1)}^*)  }} \mathcal{F}_{\textnormal{Rx},(b)}(w_{1,c}, w_{2,c}, w_{1,p}, w_{2,p}, j_1, j_2)   \Bigg)\nonumber \\
				& & \hspace{1cm} \cup \quad \mathcal{F}_{\textnormal{Rx},(b)}\left(W_{1,c,(b-1)}, {W}_{2,c,(b-1)}, W_{1,p,(b)}, W_{2,p,(b)}, J_{1,b-1}^*, J_{2,(b-1)}^*\right)  
				\nonumber\\
				&&\hspace{9cm} \Bigg| \; \bigcup_{b'=1}^{B+1} \left\{ \bar{ \mathcal{E}}_{\textnormal{Tx},1,(b')}, \; \bar{ \mathcal{E}}_{\textnormal{Tx},2,(b')} \right\}\
				\Bigg)   \IEEEeqnarraynumspace\\
				& \leq & 
				\sum_{\substack{(w_{1,c}, w_{2,c}, w_{1,p}, w_{2,p}, j_1, j_2)   \neq \\(W_{1,c,(b-1)}, {W}_{2,c,(b-1)}, W_{1,p,(b)}, \\ \hspace{1cm}W_{2,p,(b)}, J_{1,b-1}^*, J_{2,(b-1)}^*)}  }
				\Pr \Bigg(  \mathcal{F}_{\textnormal{Rx},(b)}(w_{1,c}, w_{2,c}, w_{1,p}, w_{2,p}, j_1, j_2)   \Bigg)  \; \Bigg| \; \bigcup_{b'=1}^{B+1} \left\{ \bar{ \mathcal{E}}_{\textnormal{Tx},1,(b')}, \; \bar{ \mathcal{E}}_{\textnormal{Tx},2,(b')} \right\}\
				\Bigg) 
				\nonumber \\
				&& \hspace{0cm} + \quad \Pr\Bigg( \mathcal{F}_{\textnormal{Rx},(b)}\left(W_{1,c,(b-1)}, {W}_{2,c,(b-1)}, W_{1,p,(b)}, W_{2,p,(b)}, J_{1,b-1}^*, J_{2,(b-1)}^*\right)   \; 
				\Bigg| 
				\nonumber\\&&\hspace{10cm}
				\; \bigcup_{b'=1}^{B+1} \left\{ \bar{ \mathcal{E}}_{\textnormal{Tx},1,(b')}, \; \bar{ \mathcal{E}}_{\textnormal{Tx},2,(b')} \right\}\Bigg)
			\end{IEEEeqnarray}
		\end{subequations}
		where the inequalities hold by the union bound.
		
		By the event in the conditioning and the way we construct the codebooks,  and  by the weak law of large numbers and the Covering Lemma, both summands tend to 0 as $N\to \infty$ if \eqref{eq:subrates} hold.

		The scheme satisfies the distortion constraints \eqref{eq:asymptotics_dis} because of \eqref{Th:distortion:MAC} and by the weak law of large numbers.
	}
	
	\lo{		\section{Fourier-Motzkin Elimination}\label{App:FME:MAC}
		We apply the Fourier-Motzkin Elimination Algorithm to show that Constraints~\eqref{eq:subrates} are equivalent to the constraints in Theorem~\ref{Th:achievability:MAC}.  For ease of notation,  define
		\begin{subequations}
			\mw{\begin{IEEEeqnarray}{rCl}
					I_0 & :=& I( V_1; X_1X_2Y \mid \underline{U}) 	+ I( V_2; X_1X_2YV_1 \mid \underline{U})	\\
					I_1 &:=&  I(V_1;X_1Z_1\mid \underline{U}) \\
					I_2 &:=& I(V_2;X_2Z_2\mid \underline{U}) \\
					I_3&:=&I(U_1;X_2Z_2\mid U_0U_2) \\
					I_4 &:=& I(U_2;X_1Z_1\mid U_0U_1)\\
					I_5&:=&I(V_1;X_2Z_2\mid \underline{U})\\
					I_6 &:=& I(V_2;X_1Z_1\mid \underline{U})\\
					I_7 &:=& I(X_1 X_2; YV_1V_2\mid  \underline{U})\\
					I_8 &:=& I(X_1 ; YV_1V_2\mid \underline{U} X_2 )\\
					I_9  &:=&I(X_2 ; YV_1V_2\mid \underline{U}X_1)
					\\
					I_{10}&:=& I(X_1; Y \mid U_0 X_2 ) 	\\	
					I_{11}&:=&I(X_2; Y \mid U_0 X_1) 		\\	
					I_{12}&:=&
					I(X_1X_2; Y \mid U_0 U_2 ) \\
					I_{13}&:=&
					I(X_1X_2; Y \mid U_0 U_1) 		\\
					I_{14}&:=&
					I(X_1X_2; Y \mid U_0  ) \\
					I_{15}&:=&I(X_1X_2; Y ).
			\end{IEEEeqnarray}}
		\end{subequations}
		Setting $R_{k,c}=R_k-R_{k,p}$, which is obtained from \eqref{eq:d1}, with above definitions we can rewrite Constraints \eqref{eq:subrates} as: 
		\begin{subequations}\begin{IEEEeqnarray}{rCl}
				R_{1,v} &>& I_1 \label{eq:R1vfme}\\
				R_{2,v}&>&I_2\label{eq:R2v}\\ 
				\mw{R_{2,v}+}R_{1}-R_{1,p}&<&I_2+ I_3\label{eq:R1c}\\
				\mw{R_{1,v}+}R_{2}-R_{2,p}&<&  I_1+I_4
				\label{eq:R2cfme}\\
				R_{1,v}+\mw{R_{2,v}+}R_{1}-R_{1,p}&<&I_2+ I_3+I_5
				\label{eq:R1vR1c}
				\\
				\mw{R_{1,v}+}R_{2,v}+R_{2}-R_{2,p}&<& I_1+I_4+I_6
				\label{eq:R2vR2c}\\
				R_{1,p}+R_{2,p} &<& I_7
				\label{eq:R1pR2p_FME}
				\\
				R_{1,p} &< & I_8
				\\
				R_{2,p}& < & I_9
				\\
				R_{1,v}+R_{1,p}&<& I_{10}+I_0		
				\\	
				R_{2,v}+R_{2,p}&<& I_{11}+I_0
				\\	
				R_{1,v}+R_{1,p}+R_{2,p}&<&
				I_{12}+I_0\\ 
				R_{2,v}+R_{1,p}+R_{2,p}&<& I_{13}+I_0
				\\
				R_{1,v} +R_{1,p}+ R_{2,v}+R_{2,p}&<&
				I_{14}+I_0
				\\
				R_{1,v}+R_1 +R_{2,v}+R_2 & <&  I_{15}+I_0.\label{FME1}
			\end{IEEEeqnarray}
		\end{subequations}
		In a next step we eliminate the variables $R_{1,v}$ and $R_{2,v}$ to obtain:
		\begin{subequations}\begin{IEEEeqnarray}{rCl}
				R_1-R_{1,p} &<&I_3\label{eq:d2}\\
				R_2-R_{2,p}&<&I_4\label{eq:d3}\\ 
				R_1-R_{1,p}&<&I_3+I_5-I_1\label{eq:d4}\\
				R_2-R_{2,p}&<& I_4+I_6-I_2\label{eq:d5}\\
				R_{1,p} &<&\min\{  I_8,  I_{10}+I_0-I_1\}\\
				R_{2,p} &<&\min\{ I_9, I_{11}+I_{0}-I_2\}\\
				R_{1,p}+R_{2,p}&<&\min\{I_7, I_{12}+I_0-I_1,\nonumber \\
				&& \qquad I_{13}+I_0-I_2, I_{14}+I_0-I_1-I_2\} \IEEEeqnarraynumspace\\
				R_{1}+R_{2}&<&I_{15}+I_0-I_1-I_2
			\end{IEEEeqnarray}
		\end{subequations}
		Notice that $I_{1} \geq I_5$ and $I_{2} \geq I_6$ because $V_1-(Z_1X_1\underline{U})-(X_2Z_2)$ form a Markov chain, and thus Constraints \eqref{eq:d2} and \eqref{eq:d3} are inactive in view of Constraints \eqref{eq:d4} and \eqref{eq:d5}. We  thus neglect \eqref{eq:d2} and \eqref{eq:d3} in the following. Eliminating next variable $R_{1,p}$, where we take into account the nonnegativity of $R_{1,p}$ and $R_{1}-R_{1,p}$, we obtain: 
		\begin{subequations}\begin{IEEEeqnarray}{rCl}
				R_1 &<&I_3 +I_5-I_1+\min\{  I_8,  I_{10}+I_0-I_1\}\IEEEeqnarraynumspace\\
				R_1 +R_{2,p} &<&I_3+I_5-I_1 + \min\{I_7, I_{12}+I_0-I_1,\nonumber \\
				&& \qquad I_{13}+I_0-I_2, I_{14}+I_0-I_1-I_2\} \IEEEeqnarraynumspace\\
				R_2-R_{2,p}&<& I_4+I_6-I_2\\
				R_{2,p} &<&\min\{ I_9, I_{11}+I_{0}-I_2\} \label{eq:f1}\\
				R_{2,p}&<&\min\{I_7, I_{12}+I_0-I_1,\nonumber \\
				&& \qquad I_{13}+I_0-I_2, I_{14}+I_0-I_1-I_2\}\label{eq:f2} \IEEEeqnarraynumspace\\
				R_{1}+R_{2}&<&I_{15}+I_0-I_1-I_2
			\end{IEEEeqnarray}
			and 
			\begin{IEEEeqnarray}{rCl}
				I_3+I_5&>& I_1\\
				I_{10}+I_0 &>& I_1.
			\end{IEEEeqnarray}
		\end{subequations}
		Notice that $I_7 > I_9$ and $I_{13}>I_{11}$ and therefore the two Constraints \eqref{eq:f1} and \eqref{eq:f2} combine to 
		\begin{IEEEeqnarray}{rCl}
			R_{2,p} &<&\min\{ I_9, I_{11}+I_{0}-I_2, \nonumber \\
			&& \qquad I_{12}+I_0-I_1,I_{14}+I_0-I_1-I_2\}.
		\end{IEEEeqnarray}
		Eliminating finally 
		$R_{2,p}$ (while taking into account the nonnegativity of $R_{2,p}$ and $R_{2}-R_{2,p}$) results in: 
		\begin{subequations}
			\begin{IEEEeqnarray}{rCl}
				R_1&<&I_3+I_5-I_1+\min\{ I_8,\;  I_{10}+I_0-I_1\} \label{eq:fff}\IEEEeqnarraynumspace \\
				R_1 &<&I_3+I_5-I_1 + \min\{I_7, I_{12}+I_0-I_1,\nonumber \\
				&& \qquad I_{13}+I_0-I_2, I_{14}+I_0-I_1-I_2\}\label{eq:fffDk}
				\\
				%
				R_2&<& I_4+I_6-I_2+\min\{I_9, \;  I_{11}+I_{0}-I_2 \nonumber \\
				&& \qquad I_{12}+I_0-I_1,I_{14}+I_0-I_1-I_2\}\\
				R_1+R_2&<& I_4+I_6- I_2+I_3+I_5- I_1\nonumber\\
				&&\hspace{0cm}+	\min\{I_7, \; I_{12}+I_0-I_1,
				\nonumber\\&&\hspace{1cm}I_{13}+I_0-I_2,I_{14}+I_0-I_1-I_2
				\} 	\\
				R_{1}+R_{2}&<&I_{15}+I_0-I_1-I_2
			\end{IEEEeqnarray}
			and 
			\begin{IEEEeqnarray}{rCl}
				I_3+I_5 &>& I_1\\
				I_4 +I_6 & >& I_2\\
				I_{14}+I_0& >& I_1 +I_2\\
				I_{10}+I_0 &>& I_1\label{eq:d0}\\
				I_{11} +I_0 & > & I_2\\
				I_{12} +I_0 & > & I_1.\label{eq:d}
			\end{IEEEeqnarray}
		\end{subequations}
		Notice that $I_{12}>I_{10}$ and thus \eqref{eq:d} is obsolete \mw{in view of \eqref{eq:d0}. Moreover, since} also $I_7>I_8$, Constraints \eqref{eq:fff} and \eqref{eq:fffDk} combine to 
		\begin{IEEEeqnarray}{rCl}
			R_1&<&I_3+I_5-I_1+\min\{ I_8,\;  I_{10}+I_0-I_1,\nonumber \\
			&& \qquad I_{13}+I_0-I_2, I_{14}+I_0-I_1-I_2\} .\IEEEeqnarraynumspace
		\end{IEEEeqnarray}
		The final expression is thus given by constraints: 
		\begin{subequations}
			\begin{IEEEeqnarray}{rCl}
				R_1&<&I_3+I_5-I_1+\min\{ I_8,\;  I_{10}+I_0-I_1\nonumber \\
				&& \qquad I_{13}+I_0-I_2, I_{14}+I_0-I_1-I_2\} \IEEEeqnarraynumspace\\
				R_2&<& I_4+I_6-I_2+\min\{I_9, \;  I_{11}+I_{0}-I_2 \nonumber \\
				&& \qquad I_{12}+I_0-I_1,I_{14}+I_0-I_1-I_2\}\\
				R_1+R_2&<& I_4+I_6- I_2+I_3+I_5- I_1\nonumber\\
				&&\hspace{0cm}+	\min\{I_7, \; I_{12}+I_0-I_1,
				\nonumber\\&&\hspace{1cm}I_{13}+I_0-I_2,I_{14}+I_0-I_1-I_2
				\} 	\\
				R_{1}+R_{2}&<&I_{15}+I_0-I_1-I_2
			\end{IEEEeqnarray}
			and 
			\begin{IEEEeqnarray}{rCl}
				I_3+I_5 &>& I_1\\
				I_4 +I_6 & >& I_2\\
				I_{14}+I_0& >& I_1 +I_2\\
				I_{10}+I_0 &>& I_1\\
				I_{11} +I_0 & > & I_2.
			\end{IEEEeqnarray}
		\end{subequations}
	}
	
	\bibliographystyle{IEEEtran}
	\bibliography{JRC,main_v2}	
	
\end{document}